\documentclass[a4paper,journal]{IEEEtran}
\usepackage[colorlinks,
            linkcolor=blue,
            anchorcolor=blue,
            citecolor=blue]{hyperref}
\usepackage{amsmath,amsfonts}
\usepackage{amsthm,amssymb,lipsum}
\usepackage{xcolor}

\usepackage{mathrsfs}
\usepackage[linesnumbered,ruled,vlined]{algorithm2e}
\SetKwInput{KwInput}{Input}                
\SetKwInput{KwOutput}{Output}              
\SetKwBlock{KwInitialize}{Initialize}{}

\usepackage{array}
\usepackage[caption=false,font=normalsize,labelfont=rm,textfont=rm]{subfig}
\usepackage{algpseudocode}
\usepackage{textcomp}
\usepackage{stfloats}
\usepackage{booktabs}
\usepackage{url}
\usepackage{verbatim}
\usepackage{graphicx}
\usepackage{float}
\usepackage{cite}
\usepackage{enumitem}
\usepackage{tabularx}
\usepackage{array}
\usepackage[numbers, square]{natbib}
\usepackage{flushend}
\usepackage{geometry}
\begin{document}

\title{Cross-Layer Encrypted Semantic Communication Framework for Panoramic Video Transmission}

\author{
    Haixiao Gao, \IEEEmembership{Student Member, IEEE}, 
    Mengying Sun, \IEEEmembership{Member, IEEE},
    Xiaodong Xu, \IEEEmembership{Senior Member, IEEE}, 
    Bingxuan Xu, \IEEEmembership{Student Member, IEEE},
    Shujun Han, Bizhu Wang, Sheng Jiang, Chen Dong, \IEEEmembership{Member, IEEE}, \\
    Ping Zhang, \IEEEmembership{Fellow, IEEE}

\thanks{This paper is supported by the National Key R\&D Program of China No. 2020YFB1806900, in part by the China Postdoctoral Science Foundation under Grant 2023M740341, and in part by the Beijing Natural Science Foundation No. L232051. (\emph{Corresponding author: Xiaodong Xu.})

Haixiao Gao, Mengying Sun, Bingxuan Xu, Shujun Han, Bizhu Wang, Sheng Jiang, and Chen Dong are with the State Key Laboratory of Networking and Switching Technology, Beijing University of Posts and Telecommunications, Beijing 100876, China (e-mail: haixiao@bupt.edu.cn, smy\_bupt@bupt.edu.cn, xubingxuan@bupt.edu.cn, hanshujun@bupt.edu.cn, wangbizhu\_7@bupt.edu.cn, shengjiang@bupt.edu.cn, dongchen@bupt.edu.cn).

Xiaodong Xu and Ping Zhang are with the State Key Laboratory of Networking and Switching Technology, Beijing University of Posts and Telecommunications, Beijing 100876, China, and also with the Department of Broadband Communication, Peng Cheng Laboratory, Shenzhen 518066, Guangdong, China (e-mail: xuxiaodong@bupt.edu.cn, pzhang@bupt.edu.cn).}

}

\maketitle
\begin{abstract}
In this paper, we propose a cross-layer encrypted semantic communication (CLESC) framework for panoramic video transmission, incorporating feature extraction, encoding, encryption, cyclic redundancy check (CRC), and retransmission processes to achieve compatibility between semantic communication and traditional communication systems. Additionally, we propose an adaptive cross-layer transmission mechanism that dynamically adjusts CRC, channel coding, and retransmission schemes based on the importance of semantic information. This ensures that important information is prioritized under poor transmission conditions. To verify the aforementioned framework, we also design an end-to-end adaptive panoramic video semantic transmission (APVST) network that leverages a deep joint source-channel coding (Deep JSCC) structure and attention mechanism, integrated with a latitude adaptive module that facilitates adaptive semantic feature extraction and variable-length encoding of panoramic videos. The proposed CLESC is also applicable to the transmission of other modal data. Simulation results demonstrate that the proposed CLESC effectively achieves compatibility and adaptation between semantic communication and traditional communication systems, improving both transmission efficiency and channel adaptability. Compared to traditional cross-layer transmission schemes, the CLESC framework can reduce bandwidth consumption by 85\% while showing significant advantages under low signal-to-noise ratio (SNR) conditions.
\end{abstract}

\vspace{-2mm}
\section{Introduction}


\IEEEPARstart{W}{ith} the rapid development of information technology and mobile devices, global data traffic is experiencing unprecedented growth. It is projected that by 2027, data traffic on smart devices will reach 41 GB per month, nearly four times the level in 2021 \cite{mobile_data_traffic}. Against this backdrop, the upcoming 6G era must be prepared to support a wider and more diverse range of application scenarios to meet future challenges. For example, immersive traffic information scenarios require systems to provide real-time, panoramic traffic information to ensure that individuals and vehicles receive high-quality and responsive data. At the same time, ultra-large-scale connectivity scenarios anticipate that billions of devices will be connected and communicating simultaneously in future urban environments. This raises higher demands for data processing capabilities and intelligent network management \cite{itu}, \cite{6g_1}, \cite{immersive_communication}.

To meet these increasingly severe demands, new communication technologies must be developed. Semantic communication, as one of the key technologies for 6G, has attracted widespread attention \cite{semantic_communication1}, \cite{semantic_communication2}. It significantly reduces the need for data transmission by extracting and compressing information at the semantic level while ensuring the precision and relevance of the information conveyed. Compared to traditional bit-level error-free transmission, semantic communication aims to pursue higher ``semantic fidelity'' rather than error-free transmission of symbols, and effectively overcomes the ``cliff effect'' caused by declines in signal-to-noise ratios (SNR) \cite{semantic_cliff_effect}. Therefore, as an emerging communication paradigm, semantic communication not only addresses the challenges of large-scale data transmission but also lays the foundation for an efficient and intelligent future communication network.

\begin{figure}[tbp]
\centering
\includegraphics[width=85mm]{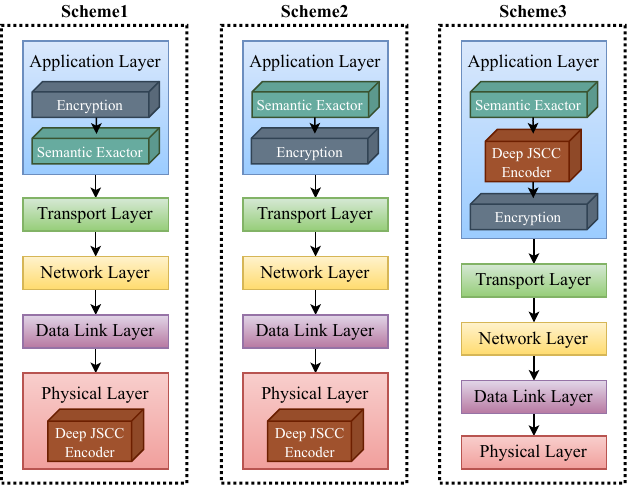}
\caption{The schemes of semantic communication and traditional mobile communication framework compatibility and adaptation.}
\vspace{-2mm}
\label{differernt cross-layer schemes}
\end{figure}

Traditional mobile communication systems adopt a five-layer TCP/IP protocol architecture, including the physical layer, data link layer, network layer, transport layer, and application layer \cite{TCP_IP}. This structure ensures the reliability, efficiency, and standardization of communication services, enabling seamless connectivity with the Internet and other TCP/IP-based networks. However, the design of semantic communication frameworks currently focuses on different transmission information modalities (such as image, video, text, speech, etc.), with targeted designs for semantic extraction networks and deep joint source-channel coding (JSCC) \cite{ntscc}, \cite{DVST}, \cite{text_semantic}, \cite{speech_semantic}. The processing of semantic information is concentrated in the data link layer or physical layer without considering the impact on the transmission performance of other layers. To ensure compatibility with traditional mobile communication systems and coordination and information sharing between layers, as shown in Fig.~\ref{differernt cross-layer schemes}, we present several compatible adaptation schemes:
\begin{itemize}
    \item \textit{Scheme 1}: Encryption and feature extraction processes are performed at the physical layer with Deep JSCC encoding, where encryption precedes feature extraction. However, this scheme disrupts source correlation and ineffective semantic information extraction.
    \item \textit{Scheme 2}: Similar to \textit{Scheme 1}, the encryption process is executed after semantic extraction. However, this scheme destroys the correlation of semantic information, reducing the efficiency of Deep JSCC encoding.
    \item \textit{Scheme 3}: Feature extraction, Deep JSCC encoding, and encryption are performed at the physical layer. However, how other layers recognize the semantics of information remains an unresolved issue.
\end{itemize}

In light of this, based on the priority of semantic information, our study proposes a cross-layer encrypted semantic communication (CLESC) framework, aiming to achieve efficient transmission while ensuring seamless compatibility with traditional mobile communication systems. Specifically, the physical and data link layers within our framework dynamically adjust resource allocation and error control strategies based on the importance of semantic information (hereinafter referred to as semantic importance). In addition, the application layer is restructured to better identify the importance and priority of information, integrating closely with the underlying transport characteristics. This ensures that important information is prioritized for transmission and processing, even under poor transmission conditions. These designs not only enhance the efficiency of communication but also improve the system's adaptability and responsiveness in dynamic environments.

Furthermore, we use panoramic video transmission as an example to demonstrate the superiority of the proposed cross-layer encrypted semantic communication framework. Panoramic video, as a crucial component of immersive communication within virtual reality (VR) and augmented reality (AR), poses significant demands on network performance due to its massive data volume \cite{apvst}, \cite{rmsa_apvst}. This first necessitates ensuring the efficiency of end-to-end panoramic video semantic transmission. In this paper, to transmit the panoramic video, equirectangular projection (ERP) is first applied to project spherical video on a flat surface. However, due to the characteristics of the ERP \cite{End_to_End_Optimized_360_Image_Compression}, there would be a large information redundancy if panoramic videos are transmitted directly by current semantic video transmission networks \cite{DCVC}, \cite{DVST}. At the same time, it will cause a decrease in the panoramic video evaluation metrics weighted-to-spherically-uniform peak signal-to-noise ratio (WS-PSNR) and weighted-to-spherically-uniform structural similarity (WS-SSIM) which mean the quality of immersive experience \cite{WS_PSNR}, \cite{WS_SSIM}.

To enhance the transmission efficiency of panoramic video and provide a superior immersive experience quality, while also facilitating cross-layer transmission and adaptation of semantic features, a cross-layer transmission framework for panoramic video based on semantic communication needs to be designed. The contributions of this paper can be summarized as follows:
\begin{itemize}
    \item We propose the CLESC framework, a novel approach that integrates semantic communication with traditional communication systems. The CLESC framework integrates feature extraction, encoding, encryption, cyclic redundancy check (CRC), and retransmission strategies. By identifying the semantic importance, it enables underlying layers to adaptively adjust transmission strategies, ensuring the system's transmission efficiency in environments with low SNRs and limited resources. To our knowledge, this is the first to achieve compatibility between semantic communication and traditional communication systems.
    \item We propose an adaptive cross-layer transmission mechanism that dynamically adjusts CRC, channel encoding, and retransmission schemes based on semantic importance, prioritizing the transmission of important information. This mechanism enhances the adaptive capacity and efficiency of cross-layer information transmission across diverse environments.
    \item We design a semantic transmitter and receiver for adaptive panoramic video semantic transmission (APVST) to verify the efficiency of the CLESC framework. Based on the Deep JSCC structure and attention mechanisms, APVST extracts semantic features from panoramic frames and implements variable-length encoding of information. Additionally, APVST utilizes an entropy model to mark the semantic importance level based on the entropy, ensuring priority transmission of crucial data. Furthermore, a weight attention (WA) module and a latitude adaptive module are introduced to enhance the quality and efficiency of panoramic video transmission.
    \item The simulation results demonstrate that the proposed CLESC achieves compatibility and adaptation between semantic communication and traditional communication systems, showing improvements in both transmission efficiency and noise resistance. Compared to traditional communication schemes, APVST reduces bandwidth consumption by 85\% and exhibits significant advantages at low SNR in cross-layer transmission.
\end{itemize}

The rest of this paper is organized as follows. The related work is introduced in Section \ref{related work}. The system model and optimization goal are described in Section \ref{system model}. The proposed adaptive cross-layer transmission mechanism is described in Section \ref{adaptive cross-layer transmission mechanism}. Section \ref{enabled modules and internal structure} introduces the enabled modules and internal structures of APVST. Experimental results are analyzed in Section \ref{section experiment}. Conclusion is finally drawn in Section \ref{conclusion}.

\section{Related Work} \label{related work}
This section provides a brief review of the related works on semantic communication, cross-layer design, and panoramic video transmission.
\subsection{Semantic Communication}

As artificial intelligence (AI) and communication technologies converge, semantic communication has emerged as a new paradigm. Compared to traditional communication approaches, it not only achieves extreme compression of information but also ensures the quality of communication under poor channel conditions, thereby enhancing the transmission efficiency and robustness of communication systems \cite{semantic_communication3}, \cite{semantic_communication4}, \cite{semantic_communication5}. However, the current preprocessing of semantic information primarily focuses on the data link layer or physical layer. For the data link layer, Hu \emph{et al}. proposed semantic-aware hybrid automatic repeat request (SemHARQ) \cite{semharq}, which utilizes a feature distortion evaluation network and a feature importance ranking network to assess the distortion degree and importance of features, thereby avoiding unnecessary retransmissions and saving transmission bandwidth. Jiang \emph{et al}. combined semantic coding (SC) with reed-solomon (RS) channel coding and hybrid automatic repeat request (HARQ) \cite{sc_rs_harq}, enhancing system performance through a transformer-based JSCC approach. For the physical layer, T.-Y \emph{et al}. proposed DeepJSCC-Q \cite{deepjscc_q}, a novel DeepJSCC scheme for wireless image transmission that uses a discrete channel input constellation, achieving robust performance and maintaining image quality in varying channel conditions without the typical cliff-effect seen in traditional methods. Bo \emph{et al}. proposed a joint coding-modulation (JCM) framework for digital semantic communications \cite{jcm}, which learns the transition probabilities from source data to discrete constellation symbols and can align the derived modulation strategy with the operating channel conditions.

However, both approaches overlook the impacts of encryption and other layers in the TCP/IP model on transmission performance, reducing the credibility of the analysis. This necessitates the implementation of cross-layer design in semantic communications within communication systems, ensuring the compatibility of semantic models with communication devices and safeguarding the efficiency of semantic information in cross-layer transmission.

\subsection{Cross-Layer Design}
Traditional network design often adheres to a layered principle, where each layer handles its specific functions. While this strict layered structure works well in wired networks, it often fails to efficiently address issues in wireless networks due to the highly dynamic nature of wireless channels and resource limitations \cite{cross_layer_survey1}, \cite{cross_layer_survey2}. Consequently, extensive research into cross-layer design for wireless communications has been undertaken, focusing on three primary objectives: security, quality of service (QoS), and mobility. For security, Chuang \emph{et al}. proposed the cross-layer design network security management (CLDNSM) \cite{CLDNSM}, which aggregates system information from various layers to achieve optimal security settings and safeguard the system. Regarding QoS, Miyoshi \emph{et al}. proposed a forward error correction mechanism that integrates forward error correction with explicit loss notifications \cite{FEC}, allowing the sharing of transmission errors that occur in the data link and physical layers across different layers. For mobility, Bender \emph{et al}. utilize channel measurements, channel control, and interference suppression and mitigation to construct a bandwidth-efficient wireless data service based on code division multiple access (CDMA) and high data rate (HDR) \cite{cdma_hdr}.

With the advent of semantic communication, new challenges and demands have been placed on traditional cross-layer design. Semantic communication not only focuses on the integrity of data but also emphasizes the relevance and meaning of the information content in relation to the receiver's task. This necessitates that the network prioritizes and transmits information that is critical to the end task. Therefore, cross-layer design needs to integrate functions from the physical layer to the application layer to support the efficient processing of semantic information.

\subsection{Panoramic Video Transmission}
Traditional panoramic video transmission schemes employ bitrate selection, resource allocation, and other strategies to achieve low-latency and high transmission performance metrics \cite{transcoding_enabled}, \cite{vr_multicast}, \cite{vr_mobile_edge}. In the context of multiple-input multiple-output (MIMO) and orthogonal frequency-division multiple access (OFDMA), Guo \emph{et al}. constructed a MIMO-OFDMA system for adaptive panoramic video transmission \cite{OFDMA_VR}. This system is capable of adaptively selecting the quality of tiles, transmission power, and beamforming based on the current environment. Using non-orthogonal multiple access (NOMA), Li \emph{et al}. proposed a NOMA-assisted VR content transmission scheme aimed at minimizing the cost-oriented towards QoS \cite{NOMA_VR}. As demands continue to rise, these traditional transmission schemes have become inadequate.

In order to achieve efficient video transmission, there has been a lot of research on how to employ neural networks for video transmission. Li \emph{et al}. build the deep contextual video compression (DCVC) network \cite{DCVC}, which extracts rich high-dimensional contextual information by the correlation between frames to transmit videos. After that, Wang \emph{et al}. of \cite{DVST} proposed the deep video semantic transmission (DVST) network, which takes the semantic communication into account. The DVST network achieves semantic transmission of video by nonlinear transforms and deep joint source-channel coding (Deep JSCC) under the guidance of the entropy model. T.-Y \emph{et al}. proposed the DeepWiVe \cite{deepwive}, which builds on Deep JSCC and employs reinforcement learning to train a bandwidth allocation network. This optimizes the distribution strategy of limited available channel bandwidth across video frames. If panoramic videos are transmitted directly by these networks, there would be a large information redundancy and a decrease in the quality of the immersive experience. Therefore, an efficient framework for the semantic transmission of panoramic video needs to be designed.

\section{System Model and Optimization goal} \label{system model}
\begin{figure}[tbp]
\centering
\includegraphics[width=85mm]{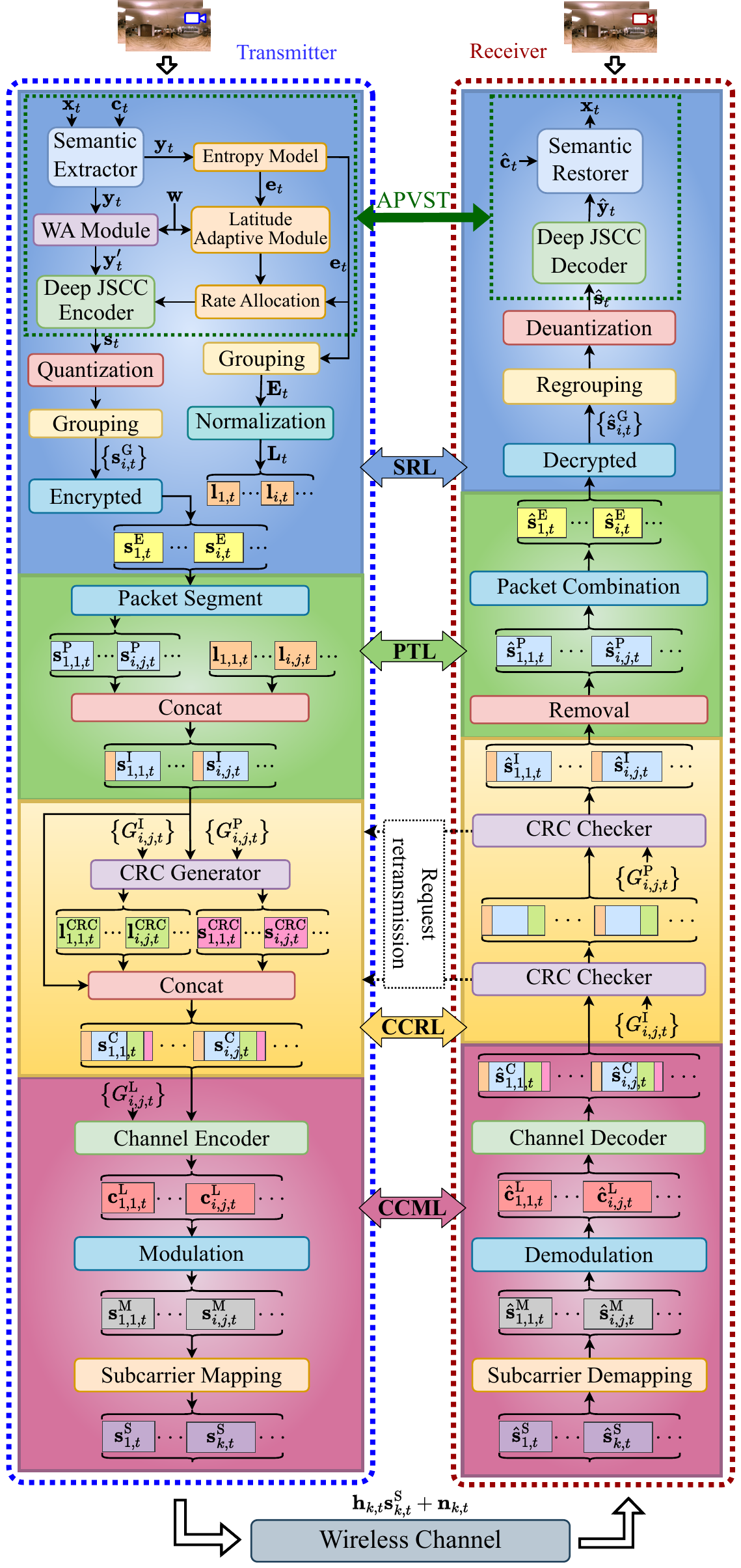}
\vspace{-3mm}
\caption{The framework of cross-layer encrypted semantic communication framework.}
\label{framework of system model}
\vspace{-3mm}
\end{figure}

We consider an end-to-end cross-layer encrypted semantic communication framework for panoramic video transmission, which adapts to the different semantic importance to ensure efficient transmission and is shown in Fig.~\ref{framework of system model}. Given that the transmitted content is video and considering the variability of the environment, a time period is divided into multiple time slots, represented by \( \mathcal{T}=\{1,2,\ldots,t,\ldots\} \). The transmission framework is structured into four distinct layers, the semantic representation layer (SRL), priority tagging layer (PTL), CRC and retransmission layer (CCRL), and channel coding and modulation layer (CCML). The definitions of some key symbols in the transmitter are shown in Table \ref{tab:symbols}.

\begin{table}[!htbp]
    \renewcommand\arraystretch{1.4}
    \centering
    \caption{Summary of Main Symbols}
    \label{tab:symbols}
    \begin{tabularx}{1\columnwidth}{|>{\centering\arraybackslash}m{0.19\linewidth}|>{\centering\arraybackslash}m{0.71\linewidth}|}
        \hline
        \textbf{Symbols} & \textbf{Definition} \\
        \hline
        \hline
        $i$, $j$, $t$ & Group index, packet index, and time slot \\
        $\mathbf{x}_t$ & Panoramic frame of $t$-th time slot \\
        $\mathbf{y}_t$, $\mathbf{s}_t$ & Feature map and codeword map of $\mathbf{x}_t$ \\
        $\mathbf{c}_t$ & Context between $\mathbf{x}_t$ and $\mathbf{x}_{t-1}$ \\
        $\mathbf{e}_t$ & Entropy map of $\mathbf{y}_t$ \\
        $\mathbf{e}^\text{G}_{i,t}$ & Grouped entropy map of $i$-th group  \\
        $\mathbf{s}^\text{G}_{i,t}$ & Grouped codeword map of $i$-th group \\
        $\mathbf{s}^\text{E}_{i,t}$ & Encrypted codeword sequence of $i$-th group \\
        $\mathbf{l}_{i,j,t}$ & Importance indication of $i$-th group and $j$-th packet \\
        $\mathbf{s}^\text{P}_{i,j,t}$ & Encrypted $j$-th packet of $i$-th group \\
        $\mathbf{s}^\text{I}_{i,j,t}$ & Concatenation codeword of $\mathbf{s}^\text{P}_{i,j,t}$ and $\mathbf{l}_{i,j,t}$ \\
        $\mathbf{s}^\text{CRC}_{i,j,t}$, $\mathbf{l}^\text{CRC}_{i,j,t}$ & CRC bits of $\mathbf{s}^\text{P}_{i,j,t}$ and $\mathbf{l}_{i,j,t}$ \\
        $\mathbf{s}^\text{C}_{i,j,t}$ & Concatenation of $\mathbf{s}^\text{I}_{i,j,t}$, $\mathbf{s}^\text{CRC}_{i,j,t}$, and $\mathbf{l}^\text{CRC}_{i,j,t}$  \\
        $\mathbf{c}^\text{L}_{i,j,t}$ & Channel encoded codeword of $j$-th packet in $i$-th group \\
        $\mathbf{s}^\text{M}_{i,j,t}$ & Modulated symbol of $j$-th packet in $i$-th group \\
        $\mathbf{s}^\text{S}_{k,t}$ & Channel transmission symbol of $k$-th subcarrier  \\
        $\mathbf{h}_{k,t}$ & Channel matrix of $k$-th subcarrier \\
        $G^\text{L}_{i,j,t}$ & Generating matrix of $j$-th packet in $i$-th group  \\
        $S^\text{P}_{i,j,t}\left(x\right)$ & Polynomial forms of $\mathbf{s}^\text{P}_{i,j,t}$ \\
        $L_{i,j,t}\left(x\right)$ & Polynomial forms of $\mathbf{l}_{i,j,t}$ \\
        $S^\text{CRC}_{i,j,t}\left(x\right)$ & Polynomial forms of $\mathbf{s}^\text{CRC}_{i,j,t}$ \\
        $L^\text{CRC}_{i,j,t}\left(x\right)$ & Polynomial forms of $\mathbf{l}^\text{CRC}_{i,j,t}$ \\
        $G^\text{P}_{i,j,t}\left(x\right)$ & CRC generating polynomial of $\mathbf{s}^\text{P}_{i,j,t}$ \\
        $G^\text{I}_{i,j,t}\left(x\right)$ & CRC generating polynomial of $\mathbf{l}_{i,j,t}$ \\
        $\mathbf{w}$, $\mathbf{M}_t$ & Weight map and weighted spatial attention map \\
        $\boldsymbol{\omega}_t$, $\boldsymbol{\eta}_t$& Adaptive weight feature map and weight factor map \\
        $H$, $W$ & Height and width of panoramic frame \\
        $m_\text{H}$, $m_\text{W}$ & Granularity of the grouping in terms of $H$ and $W$ \\
        $n_{i,t}$ & Number of packets of $i$-th group \\
        \hline
    \end{tabularx}
\end{table}

In this framework, let $\mathbf{x}_t$ represent the current panoramic video frame to be transmitted. At the transmitter's semantic representation layer, $\mathbf{x}_t$ is processed through the semantic transmitter and encryption modules to convert it into a semantic codeword sequence. Each group of these codewords is assigned an importance indication, indicated by $\mathbf{l}_{i,t}$. In the priority tagging layer, this sequence is segmented into packets, and $\mathbf{l}_{i,t}$ is included in the header of each packet. The CRC and retransmission layer is to implement varying levels of CRC and retransmission strategies based on $\mathbf{l}_{i,t}$. To prevent the ``avalanche effect'' caused by encryption, the channel coding and modulation layer incorporates redundancy to ensure the reliability of information transmission. At the receiver, the received information undergoes a series of demodulation.

\subsection{The Modules and Process at Transmitter}
In this section, the functions of the different layers during the transmitting process of panoramic videos are described.
\subsubsection{SRL}
Through the SRL, the semantic extraction and Deep JSCC encoding are first performed on the panoramic frame. We refer to the semantic model in SRL of transmitter and receiver together as APVST. Referring to \cite{DVST}, \cite{apvst}, based on the rich correlation information between current and reference panoramic frames, the motion link of APVST estimates the motion information and transmits it to the receiver. By context generator similar to in \cite{apvst}, the context ${\hat{\mathbf{c}}}_t$ in the transmitter and $\hat{\mathbf{c}}_t$ in the receiver are obtained, respectively. In the primary link, based on the context, the APVST adaptively converts the source frames into potential representations in the semantic feature domain by semantic extractor. To enable wireless transmission, the semantic information is encoded at variable length by the Deep JSCC encoder, while the code rate is jointly guided by the learnable latitude adaptive module and entropy model. Due to the similarity between the primary link and the motion link, the transmission process of panoramic video on the primary link will be introduced in detail.

As shown in Fig.~\ref{framework of system model}, the APVST automatically learns the correlation between the current frame $\mathbf{x}_t$ and $\mathbf{c}_t$ to remove the information redundancy. The semantic feature map $\mathbf{y}_t$ is extracted on the basis of $\mathbf{c}_t$ and encoded into the corresponding codeword map $\mathbf{s}_t$ by Deep JSCC encoding, which can be expressed as
\begin{equation}
\mathbf{y}_t=f_\text{SE}\left(\mathbf{x}_t\middle|\mathbf{c}_t\right)\ \text{and} \ \mathbf{s}_t=f_\text{JE}\left(\mathbf{y}_t^\prime\right),\label{eq5}
\end{equation}
where $f_\text{SE}\left(\cdot\right)$ and $f_\text{JE}\left(\cdot\right)$ denote the function of the semantic extractor and Deep JSCC encoder, respectively. As the input to the encoder, the weighted spatial attention feature $\mathbf{y}_t^\prime$ is obtained by the WA module. Simultaneously, through the entropy model, the entropy map $\mathbf{e}_t$ of the feature map $\mathbf{y}_t$ is obtained, which further indicates the amount of information that each feature point needs to transmit.

To achieve adaptive transmission of panoramic videos under various channel conditions, the codeword map $\mathbf{s}_t$ and the entropy map $\mathbf{e}_t$ are grouped, adopting different levels of transmission strategies for each group to prioritize the accurate transmission of important information. The grouping along the dimensions of height and width of the map is performed, which can be expressed as
\begin{equation}
    \mathbf{E}_t=\left\{\mathbf{e}^\text{G}_{i,t}\right\}=f_\text{G}\left(\mathbf{e}_t\right)\ \text{and} \
    \mathbf{S}_t=\left\{\mathbf{s}^\text{G}_{i,t}\right\}=f_\text{G}\left(\mathbf{s}_t\right),
\end{equation}
where $f_\text{G}\left(\cdot\right)$ denotes the function of grouping. $\mathbf{e}^\text{G}_{i,t}$ and $\mathbf{s}^\text{G}_{i,t}$ are the grouped entropy map and codeword map of group $i$ at time slot $t$. Let $m_\text{H}$ and $m_\text{W}$ denote the granularity of the grouping in terms of height and width, respectively, which corresponds to the number of divisions along the height and width. These parameters, $m_\text{H}$ and $m_\text{W}$ can be adjusted based on different requirements. Based on the entropy contained in each group, $\mathbf{E}_t$ is converted to importance indication $\mathbf{L}_t=\left\{\mathbf{l}_{i,t}\right\}$, each of which corresponds directly to the segmented codeword map $\mathbf{S}_t$. To ensure the security of the information during transmission, the vectorized $\mathbf{S}_t$ is encrypted by different encryption algorithms. This process can be expressed as
\begin{equation}
    \mathbf{s}^\text{E}_{i,t}=f_\text{E}\left(\mathbf{s}^\text{G}_{i,t}\right), \forall i \in \mathcal{I},
\end{equation}
where $f_\text{E}\left(\cdot\right)$ represents the encryption function, with each group's codeword sequence encrypted independently, without interference from others. $\mathbf{s}^\text{E}_{i,t}$ is the encrypted codeword sequence of group $i$ at time slot $t$. The index set of grouped codeword is $\mathcal{I}=\left\{i\right\}, i \in \left(1, \ldots, m_\text{H} m_\text{W}\right)$. In our encryption strategy, the information security is implemented by symmetric encryption algorithms such as the advanced encryption standard (AES) \cite{AES} and ChaCha20 \cite{chacha20} algorithms.

\subsubsection{PTL}
Referencing \cite{TCP_IP}, within the TCP/IP five-layer model, data is segmented into smaller packets to enhance the efficiency and reliability of network transmission. This approach accommodates the maximum transmission unit requirements of different networks, reducing the risk of packets being discarded due to excessive size, and also facilitates error detection and recovery. Therefore, in the PTL layer, $\mathbf{s}^\text{E}_{i,t}$ is further segmented into packets, i.e.,
\begin{equation}
    \mathbf{S}^\text{P}_{i,t}=\left\{\mathbf{s}^\text{P}_{i,j,t}\right\}=f_\text{P}\left(\mathbf{s}^\text{E}_{i,t}\right), \forall i \in \mathcal{I},
\end{equation}
where $f_\text{P}\left(\cdot\right)$ denotes the function of packet segmentation. $\mathbf{s}^\text{P}_{i,j,t}$ is the encrypted packet $j$ of group $i$ at time slot $t$.

Next, the importance indication $\mathbf{l}_{i,j,t}$ is appended to the header of data packet, where $\mathbf{l}_{i,j,t}=\mathbf{l}_{i,t}, \forall i \in \mathcal{I}$. This information is crucial for guiding the underlying processes of CRC, retransmission, and channel encoding. By integrating this approach, we ensure that these critical operations are customized based on semantic importance, thereby enhancing the efficiency and reliability of our communication system. This process demonstrates how we optimize and adjust network transmission strategies based on semantic importance, which can be expressed as
\begin{equation}
    \mathbf{s}^\text{I}_{i,j,t}=\text{Concat}\left(\mathbf{l}_{i,j,t}, \mathbf{s}^\text{P}_{i,j,t}\right), \forall (i,j) \in \mathcal{J},
\end{equation}
where $\text{Concat}\left(\cdot\right)$ denotes the function of concatenating two data packets. $\mathbf{s}^\text{I}_{i,j,t}$ is the packet $j$ of group $i$ at time slot $t$ that appended importance indication. The index set of encrypted packet is $\mathcal{J} = \{\left(i, j\right), \forall i \in \left(1,\ldots,m_\text{H} m_\text{W}\right), j \in \left(1,\ldots, n_{i,t}\right)\}$, and $n_{i,t}$ denotes the number packets of group $i$ at time slot $t$.

\subsubsection{CCRL}
To ensure the accuracy of information during transmission, a CRC bit is generated by applying a specific polynomial to the data and sent together with the data. The receiver can recalculate and compare this check. If the check does not match, it indicates that the data may have been corrupted during transmission, thus providing an effective error detection mechanism \cite{CRC}. Let $G^\text{P}_{i,j,t}\left(x\right)$ denote the CRC generating polynomial corresponding to encrypted packet $\mathbf{s}^\text{P}_{i,j,t}$, $k^\text{P}_{i,j,t}$ indicates its highest power, the process of CRC encoding can be expressed as
\begin{equation}
    S^\text{CRC}_{i,j,t}\left(x\right)= \left[S^\text{P}_{i,j,t}\left(x\right) \cdot x^{k^\text{P}_{i,j,t}}\right]_{\text{mod} \ G^\text{P}_{i,j,t}\left(x\right)}, \forall (i,j) \in \mathcal{J}, \label{packet check bits}
\end{equation}
where $\text{mod}$ represents the function for calculating the remainder of a polynomial. $S^\text{P}_{i,j,t}\left(x\right)$ and $S^\text{CRC}_{i,j,t}\left(x\right)$ refer to the polynomial form of the encrypted packet $\mathbf{s}^\text{P}_{i,j,t}$ and the CRC bits corresponding to it, respectively.

At the same time, the importance indication may also encounter errors during transmission. To ensure that the receiver can accurately identify the semantic importance of each packet, CRC is also incorporated. Let $G^\text{I}_{i,j,t}\left(x\right)$ represent the CRC generating polynomial corresponding to the importance indication $\mathbf{l}_{i,j,t}$, and $k^\text{I}_{i,j,t}$ represent its highest power. The encoding process can be expressed as
\begin{equation}
    L^\text{CRC}_{i,j,t}\left(x\right)=\left[L_{i,j,t}\left(x\right) \cdot x^{k^\text{I}_{i,j,t}}\right]_{\text{mod} \ G^\text{I}_{i,h,t}\left(x\right)}, \forall \left(i,j\right) \in \mathcal{J}, \label{improtance indication check bits}
\end{equation}
where $L_{i,j,t}\left(x\right)$ and $L^\text{CRC}_{i,j,t}\left(x\right)$ are polynomial forms of the importance indication $\mathbf{l}_{i,j,t}$ and the CRC bits corresponding to it for packet $j$, respectively. After obtaining two check bits, we sequentially append them to the end of $\mathbf{s}^\text{I}_{i,j,t}$ to form $\mathbf{s}^\text{C}_{i,j,t}$, which is expressed as
\begin{equation}
    \mathbf{s}^\text{C}_{i,j,t}=\text{Concat}\left(\mathbf{s}^\text{I}_{i,j,t}, \mathbf{s}^\text{CRC}_{i,j,t}, \mathbf{l}^\text{CRC}_{i,j,t}\right), \forall (i,j) \in \mathcal{J}, \label{concat crc to packet}
\end{equation}
where $\mathbf{s}^\text{CRC}_{i,j,t}$ and $\mathbf{l}^\text{CRC}_{i,j,t}$ are the codewords corresponding to the polynomial $S^\text{CRC}_{i,j,t}\left(x\right)$ and $L^\text{CRC}_{i,j,t}\left(x\right)$, respectively.
\subsubsection{CCML}
During transmission, if an error occurs in a group of data, the randomness of the encryption algorithm can trigger the ``avalanche effect''. As a result, a multitude of errors occur in the decrypted data. Although the generalization capability of semantic communication can mitigate this effect to some extent, under conditions of extremely low signal-to-noise ratio (SNR in negative values), this generalization capability diminishes. To achieve reliable communication at low SNR, channel coding on the codeword to be transmitted is performed in the CCML. Although this measure introduces some redundancy, it actually reduces the amount of data that needs to be transmitted compared to traditional communication methods. This reduction is due to the SRL having achieved extreme compression of the information while also adapting to channel conditions. In this case, low-density parity-check (LDPC) is employed for channel encoding \cite{LDPC}, and the process can be expressed as
\begin{equation}
    \mathbf{c}^\text{L}_{i,j,t} = G^\text{L}_{i,j,t} \cdot \mathbf{s}^\text{C}_{i,j,t}, \forall \left(i,j\right) \in \mathcal{J}, \label{channel encoding}
\end{equation}
where $\mathbf{c}^\text{L}_{i,j,t}$ and $G^\text{L}_{i,j,t}$ represent the generating matrix and the LDPC encoded codeword corresponding to $\mathbf{s}^\text{C}_{i,j,t}$, respectively. Additionally, the method of channel coding can be adapted according to different scenarios and channel conditions.

To convert codewords into physical signals suitable for the transmission medium, modulation is necessary to ensure effective transmission within the communication system. Choosing the appropriate modulation scheme under varying channel conditions can strike a balance between bit error rate and transmission speed, ensuring the reliability and efficiency of data transmission \cite{3GPP_TS_38214}. The modulation process can be expressed as
\begin{equation}
     \mathbf{s}^\text{M}_{i,j,t} = f_\text{M}\left(\mathbf{c}^\text{L}_{i,j,t}\right), \forall \left(i,j\right) \in \mathcal{J}, \label{modulation}
\end{equation}
where $f_\text{M}\left(\cdot\right)$ denotes the function of modulation. $\mathbf{s}^\text{M}_{i,j,t}$ is modulated symbol. To assess the utilization of channel resources, the channel bandwidth ratio (CBR) \cite{deepsc_f} is introduced. Let $k_t$ denote the number of modulated symbols, CBR is defined as $\frac{1}{T}\sum_{t=1}^T\frac{k_t}{H\times W\times3}$, where 3 denotes panoramic frame has three channels of RGB, and $T$ represents the time span of a group of pictures (GoP).

In this study, OFDMA is utilized for symbols transmission, where all available bandwidth is divided into multiple orthogonal subcarriers. Given the end-to-end nature of the transmission and the absence of resource contention, all data packets will be uniformly distributed across all subcarriers for transmission. Let \( K \) denote the total number of subcarriers, and \( \mathcal{K} = \{1, 2, \dots, K\} \) is the set of subcarrier indices. The transmission symbols  on the $k$-th subcarrier can be expressed as 
\begin{equation}
    \mathbf{s}^\text{S}_{k,t} = f_\text{S}\left(\mathbf{s}^\text{M}_{i,j,t}\right), \forall \left(i,j\right) \in \mathcal{J}, k \in \mathcal{K} \label{subcarrier mapping}
\end{equation}
where $f_\text{S}\left(\cdot\right)$ denote the function that mapping modulated symbols onto subcarriers. Finally, $\mathbf{s}^\text{S}_{k,t}$ are fed into the wireless channel and transmitted to the receiver.

\subsection{The Modules and Process at Receiver}
In this section, the functions of the different layers during the receiving process of panoramic videos are described in detail. In the following text, ``$\sim$'' represents the inverse operation of the original function.
\subsubsection{CCML}
During transmission, the signal on each subcarrier experiences channel fading and additive white Gaussian noise (AWGN). \( \mathbf{h}_{k,t} \) is the channel fading for each subcarrier \( k \) at time slot \( t \), and the AWGN is represented by \( \mathbf{n}_{k,t} \), where each \( \mathbf{n}_{k,t} \) follows a Gaussian distribution with zero mean and variance \( N_0 \), i.e., \( \mathbf{n}_{k,t} \sim \mathcal{N}(0, N_0) \). The received signal \( \hat{\mathbf{s}}^\text{S}_{k,t} \) on subcarrier \( k \) can be expressed as
\begin{equation}
    \hat{\mathbf{s}}^\text{S}_{k,t} = \mathbf{h}_{k,t} \mathbf{s}^\text{S}_{k,t} + \mathbf{n}_{k,t}, \quad k \in \mathcal{K}. \label{channel_model}
\end{equation}
where $\mathbf{h}_{k,t}$ is modeled as complex gain, which consists of small-scale fading and large-scale fading, i.e., $\mathbf{h}_{k,t} = \mathbf{h}^\prime_{k,t} \times 10^{-PL_t/20}$ \cite{channel_gain}. The small-scale $\mathbf{h}^\prime_{k,t}$ is modeled as Rayleigh fading, i,e. $\mathbf{h}^\prime_{k,t}\sim \mathcal{CN}\left(0, 1\right)$. The large-scale fading can be expressed as $PL_t(\text{dB})=32.4+20\log_{10}f_c+21\log_{10}d_t$, where $d_t$ and $f_c$ represent the end-to-end distance, and the carrier center frequency, respectively. Let $P$ denote the transmit power, which will be uniformly distributed to $K$ subcarriers. Furthermore, the SNR at time slot $t$ can be expressed as
\begin{equation}
    \text{SNR}_t(\text{dB})= 10\log\left(\sum^K_{k=1}\frac{P\left|\mathbf{h}_{k,t}\right|^2}{KN_0}\right) \label{cal snr}
\end{equation}

Subsequently, following subcarrier demapping, all data packets will be retrieved from their respective subcarriers. This process can be expressed as
\begin{equation}
   \hat{\mathbf{s}}^\text{M}_{i,j,t} = \tilde{f}_\text{S}\left(\hat{\mathbf{s}}^\text{S}_{k,t}\right), \forall \left(i,j\right) \in \mathcal{J}, k \in \mathcal{K}. \label{subcarrier demapping}
\end{equation}

At the same time, the demodulated codewords \( \hat{\mathbf{c}}^\text{L}_{i,j,t} \) are obtained by a decision process, which can be expressed as 
\begin{equation}
    \hat{\mathbf{c}}^\text{L}_{i,j,t} = \tilde{f}_\text{M}\left( \hat{\mathbf{s}}^\text{M}_{i,j,t} \right), \forall \left(i,j\right) \in \mathcal{J}, \label{demodulation}
\end{equation}
where \( \tilde{f}_\text{M}\left( \cdot \right) \) denotes the demodulation function, which maps the equalized symbols back to the nearest constellation points, effectively recovering the transmitted codewords.

For LDPC decoding, the belief propagation (BP) algorithm is adopted to iteratively decode the information \cite{BP_LDPC}. The decoding process of the BP algorithm begins with the initialization of messages from variable nodes to check nodes. Based on the received codeword $\hat{\mathbf{c}}^\text{L}_{i,j,t}$, the initialization process can be expressed as
\begin{equation}
\mu_{v \to c}^{(0)} = \hat{\mathbf{c}}^\text{L}_{i,j,t}\left(v\right), \quad \forall v \in V, \, c \in N(v),
\end{equation}
where $V$ represents the set of all variable nodes, and $N(v)$ is the set of check nodes connected to the variable node $v$. This process sets the initial messages that each variable node sends to its adjacent check nodes. In the next step, iterative message updating will take place, where variable nodes and check nodes exchange information to progressively refine the estimation of the codeword. Let $l$ denote the step of iterations, the update process includes:
\begin{itemize}
  \item Information update from variable nodes to check nodes:
    \begin{equation}
        \mu_{v \to c}^{(l)} = \hat{\mathbf{c}}^\text{L}_{i,j,t}\left(v\right) + \sum_{c' \in N(v) \setminus c} \mu_{c' \to v}^{(l-1)}.
    \end{equation}
  \item Information update from check nodes to variable nodes:
    \begin{equation}
        \mu_{c \to v}^{(l)} = 2 \cdot \tanh^{-1}\left[\prod_{v' \in M(c) \setminus v} \tanh\left(\frac{\mu_{v' \to c}^{(l-1)}}{2}\right)\right],
    \end{equation}
\end{itemize}
where $M(c)$ is the set of variable nodes connected to the check node $c$, and $\setminus$ represents the exclusion of the current node participating in the calculation. Let $u(x)$ represent the unit step function, and a hard decision is made based on the final messages received to obtain an estimate of the final information $\hat{\mathbf{s}}^\text{C}_{i,j,t}$ vector after a sufficient number of iterations, i.e.,
\begin{equation}
    \hat{\mathbf{s}}^\text{C}_{i,j,t}\left(v\right) = u\left(\sum_{c \in N(v)} \mu_{c \to v}^{(l)}\right), \forall \left(i,j\right) \in \mathcal{J}. \label{channel decoding}
\end{equation}

\subsubsection{CCRL}
In the receiver's CCRL, the importance indication $\hat{L}_{i,j,t}\left(x\right)$ is firstly checked. By the corresponding generating polynomial $G^\text{I}_{i,j,t}\left(x\right)$, the check bits for the received importance indication can be calculated. When these check bits $\hat{\hat{L}}^\text{CRC}_{i,j,t}\left(x\right)$ match the received importance indication check bits $\hat{L}^\text{CRC}_{i,j,t}\left(x\right)$, it indicates that the importance indication is error-free. Otherwise, it suggests an error in the importance indication, and an immediate retransmission of the data packet is requested. The process of calculating $\hat{\hat{L}}^\text{CRC}_{i,j,t}\left(x\right)$ can be expressed
\begin{equation}
    \hat{\hat{L}}^\text{CRC}_{i,j,t}\left(x\right) = \left[\hat{L}_{i,j,t}\left(x\right) \cdot x^{k^\text{I}_{i,j,t}}\right]_{\text{mod} \ G^\text{I}_{i,j,t}\left(x\right)}, \forall \left(i,j\right) \in \mathcal{J}. \label{received importance indication check bits}
\end{equation}

After checking that the importance indication has been transmitted correctly, we continue to check the received packet $\hat{S}^\text{P}_{i,j,t}\left(x\right)$ in a similar manner. The process of calculating the check bits $\hat{\hat{S}}^\text{CRC}_{i,j,t}\left(x\right)$ can be expressed as 
\begin{equation}
    \hat{\hat{S}}^\text{CRC}_{i,j,t}\left(x\right) = \left[\hat{S}^{\text{P}}_{i,j,t}\left(x\right) \cdot x^{k^\text{P}_{i,j,t}}\right]_{\text{mod} \ G^\text{P}_{i,j,t}\left(x\right)}, \forall \left(i,j\right) \in \mathcal{J}. \label{received packet data check bits}
\end{equation}

If these check bits match the received packet's check bits $\hat{S}^\text{CRC}_{i,j,t}\left(x\right)$, it indicates that the data packet is error-free. Otherwise, it means that an error has occurred in the data packet, and an immediate retransmission of the packet is requested. By this approach, $\hat{\mathbf{s}}^\text{I}_{i,j,t}$ can be obtained.

\subsubsection{PTL}
In the PTL of the receiver, the retransmission of information has been completed. At this point, the importance indication information becomes redundant and needs to be removed. Let $l_{\text{I}}$ represent the length of the importance indication $\mathbf{l}_{i,j,t}$. This process can be expressed as
\begin{equation}
    \hat{\mathbf{s}}^\text{P}_{i,j,t} = \hat{\mathbf{s}}^\text{I}_{i,j,t}[l_{\text{I}}+1:end], \forall \left(i,j\right) \in \mathcal{J}.
\end{equation}

Additionally, the data packets are combined to form the received packet codeword sequence $\hat{\mathbf{s}}^\text{E}_{i,t}$, which can be expressed as
\begin{equation}
    \hat{\mathbf{S}}^\text{P}_{i,t}=\{\hat{\mathbf{s}}^\text{P}_{i,j,t}\}\ \text{and} \ \hat{\mathbf{s}}^\text{E}_{i,t} = \tilde{f}_\text{P}\left(\hat{\mathbf{S}}^\text{P}_{i,t}\right), \forall i \in \mathcal{I}. 
\end{equation}

\subsubsection{SRL}
In the SRL of the receiver, the encrypted codeword sequence is first decrypted by the corresponding algorithm. This process may experience the ``avalanche effect'', which is why the channel coding is employed in CCML. This process can be expressed as
\begin{equation}
    \hat{\mathbf{s}}^\text{G}_{i,t} = \tilde{f}_\text{E}\left(\hat{\mathbf{s}}^\text{E}_{i,t}\right).
\end{equation}

At the same time, the decrypted data is regrouped, i.e.,
\begin{equation}
    \hat{\mathbf{S}}_t=\{\hat{\mathbf{s}}^\text{G}_{i,t}\}\ \text{and} \ \hat{\mathbf{s}}_t = \tilde{f}_\text{G}\left(\hat{\mathbf{S}}_t\right).
\end{equation}
At the semantic receiver, the received codeword map $\hat{\mathbf{s}}_t$ is converted to the recovered frame ${\hat{\mathbf{x}}}_t$ by Deep JSCC decoding and semantic recovery, which can be expressed as
\begin{equation}
    {\hat{\mathbf{y}}}_t=f_\text{JD}\left({\hat{\mathbf{s}}}_t\right)\ \text{and} \ {\hat{\mathbf{x}}}_t=f_\text{SR}\left({\hat{\mathbf{y}}}_t\middle|{\hat{\mathbf{c}}}_t\right),\label{eq6}
\end{equation}
where $f_\text{JD}\left(\cdot\right)$ and $f_\text{SR}\left(\cdot\right)$ denote the function of semantic restorer and Deep JSCC decoder, respectively. 

The APVST network is adept at autonomously learning the inter-frame correlations in panoramic videos, which is crucial for both semantic extraction and recovery. It is specifically designed to optimize key metrics like WS-PSNR and WS-SSIM during frame transmission, using a weighted spatial attention map to enhance the viewing experience. The network’s encoding and decoding phases are precisely directed by the entropy model and a latitude adaptive module, which recognizes and adjusts to the unique characteristics of panoramic videos. This enables precise, region-specific encoding, reducing information redundancy and ensuring efficient video transmission.

Throughout the process of cross-layer transmission of panoramic videos, the CLESC completes the encoding and encryption of semantic information at the application layer, ensuring secure transmission between layers. Additionally, to enable other layers to comprehend the semantic information, importance indications are embedded within the data to prioritize transmission and processing. This approach minimally achieves compatibility between semantic communication and traditional communication frameworks.

\subsection{The Optimization Goal} 
As the CLESC is an end-to-end transmission framework, there is no competition for resources among users. Therefore, the overall optimization goal of the network focuses on how to accomplish end-to-end optimization of the APVST network. 

The APVST network aims to maximize the transmission quality of frames with the minimum channel bandwidth overhead. In fact, the frames are transmitted continuously, $N$ consecutive frames are one group of pictures (GoP). In the training phase, the previously recovered panoramic frame is used as the reference frame for the current frame to achieve panoramic video transmission. Therefore, the loss function can be expressed as 
\begin{align}
    \mathcal{L} = \frac{1}{N}\sum_{t=1}^{N}\left[{D\left(\mathbf{x}_t, \hat{\mathbf{x}}_t\right) + \alpha\left(\mathcal{L}_t^{\text{enml}} + \mathcal{L}_t^{\text{en}}\right) + \beta\mathcal{L}^\prime{_t^\text{la}}}\right],\label{all loss of apvst}
\end{align}
where $D\left(\mathbf{x}_t, \hat{\mathbf{x}}_t\right)$ represents the magnitude of image distortion, which is chosen as the WMSE for WS-PSNR, and as the inverse of WS-SSIM for WS-SSIM to indicate the distortion of frames \cite{WS_PSNR}, \cite{WS_SSIM}. $\mathcal{L}^\prime{_t^\text{la}}$ denote the loss of latitude adaptive module, which is introduced in the Eq.~\eqref{loss of la}. $\alpha$ and $\beta$ denote the balance coefficients, which are constants. Let $e_t(m,n)$ and $e_t^\text{ml}(m,n)$ denote the entropy value at point $(m,n)$ corresponding to the entropy map $\mathbf{e}_t$ in the primary link and $\mathbf{e}^\text{ml}_t$ in the motion link, respectively. To minimize channel bandwidth overhead, we consider the entropy losses $\mathcal{L}_t^{\text{enml}}$ and $\mathcal{L}_t^{\text{en}}$, which can expressed as
\begin{equation}
    \mathcal{L}_t^{\text{enml}}=\sum_{m}\sum_{n}{e_t^\text{ml}(m,n)} \ \text{and} \ \mathcal{L}_t^{\text{en}}=\sum_{m}\sum_{n}{e_t(m,n)}.
\end{equation}

In section \ref{enabled modules and internal structure}, the internal structure of the APVST network will be introduced, and the optimization of the network will be completed based on this loss function.

\section{Adaptive Cross-Layer Transmission Mechanism} \label{adaptive cross-layer transmission mechanism}
Panoramic videos capture a comprehensive view of the surrounding environment, which may include repetitive and relatively unimportant elements such as the sky and grass, while also encompassing key human subjects. Although this all-encompassing perspective ensures the completeness of the scene, some content may not be essential for the overall narrative. In cases of poor channel quality, to ensure the QoS for users with limited resources, we can lower the transmission priority of unimportant content and increase the priority of crucial content. 

For this purpose, we propose an adaptive cross-layer transmission mechanism based on semantic importance. The details of this mechanism in transmitter and receiver are shown in Algorithm \ref{transmit priority update} and \ref{received priority update}, respectively. Although the transmission priority of some content is reduced, the generalization ability of the semantic model still ensures that these contents are presented to the user in the best possible quality.

Overall, the proposed mechanism operates the transmission process based on three mapping functions:
\begin{itemize}
    \item $f_{\mathbf{l} \to G^\text{P}\left(x\right)}\left(\cdot\right)$: The function that maps $\mathbf{l}_{i,j,t}$ to CRC generating polynomial $G^\text{P}_{i,j,t}\left(x\right)$. This mapping function allows the selection of appropriate CRC generating polynomials based on importance indication. Different polynomials produce varying numbers of CRC bits, higher-priority packets will have longer CRC bits to enhance error detection capabilities, while lower-priority packets will have fewer check bits.
    \item $f_{\mathbf{l} \to G^\text{L}}\left(\cdot\right)$: The function that maps $\mathbf{l}_{i,j,t}$ to channel encoding  generating matrix $G^\text{L}_{i,j,t}$. This mapping function enables the selection of different channel encoding matrices based on importance indication. These matrices vary in encoding efficiency; packets with higher priority will use matrices with lower efficiency, ensuring the accuracy of codeword transmission by incorporating redundancy. Conversely, higher encoding efficiencies are used for lower-priority packets.
    \item $f_{\mathbf{l} \to T^\text{R}}\left(\cdot\right)$: The function that maps $\mathbf{l}_{i,j,t}$ to retransmission times $T^\text{R}_{i,j,t}$. Through this mapping function, different retransmission times are assigned based on the importance indication. Higher-priority packets will undergo more retransmissions, effectively ensuring the accuracy of codeword transmission. In contrast, the retransmission count will be reduced for lower-priority packets.
\end{itemize}

\begin{algorithm}[t]  \label{transmit priority update}
    \caption{Priority-based Transmitting Pipeline}
    \KwInput{The encrypted packets set $\left\{\mathbf{s}^\text{P}_{i,j,t}\right\}$ and the importance indication set $\left\{\mathbf{l}_{i,j,t}\right\}$.}
    Initialize the mapping functions $f_{\mathbf{l} \to G^\text{P}\left(x\right)} \left(\cdot\right)$ and $f_{\mathbf{l} \to G^\text{L}}\left(\cdot\right)$. \\
    \For{$\left(i,j\right) \in \mathcal{J}$}{
        Compute the packet check bits $\mathbf{s}^\text{CRC}_{i,j,t}$ according to Eq. \eqref{packet check bits} based on the CRC generating polynomial $f_{\mathbf{l} \to G^\text{P}\left(x\right)}\left(\mathbf{l}_{i,j,t}\right)$. \\
        Compute the importance indication check bits $\mathbf{l}^\text{CRC}_{i,j,t}$ according to Eq. \eqref{improtance indication check bits}. \\
        Concatenate all codewords according to Eq. \eqref{concat crc to packet}. \\
        Compute the encoded codeword $\mathbf{c}^\text{L}_{i,j,t}$ according to Eq. \eqref{channel encoding} based on the channel encoding generating matrix $f_{\mathbf{l} \to G^\text{L}}\left(\mathbf{l}_{i,j,t}\right)$. \\
        Modulate the codewords to symbols according to Eq. \eqref{modulation}. \\
        Map symbols to subcarriers according to Eq. \eqref{subcarrier mapping}.
    }
    \KwOutput{The channel transmission symbol set $\left\{\mathbf{s}^\text{S}_{k,t}\right\}$.}
\end{algorithm}

\begin{algorithm}[t]  \label{received priority update}
    \caption{Priority-based Receiving Pipeline}
    \KwInput{The received channel transmission symbol set $\left\{\hat{\mathbf{s}}^\text{S}_{k,t}\right\}$ and the maximum retransmission times $T^\text{I}$ of importance indication.}
    Initialize the mapping function $f_{\mathbf{l} \to T^\text{R}}\left(\cdot\right)$. \\
    Demap subcarriers to obtain $\hat{\mathbf{s}}^\text{M}_{i,j,t}$ according to Eq. \eqref{subcarrier demapping}. \\
    \For{$\left(i,j\right) \in \mathcal{J}$}{
        Demodulate the codewords according to Eq. \eqref{demodulation}. \\
        Compute the decoded codeword $\hat{\mathbf{c}}^\text{L}_{i,j,t}$ according to Eq. \eqref{channel decoding} based on the corresponding channel encoding check matrix. \\
        Initialize the retransmission counter $u \leftarrow 1$, and the importance indication for the $0$-th retransmission:
        \begin{equation}
            \setlength\abovedisplayskip{1pt}
            \setlength\belowdisplayskip{1pt}
            \hat{L}_{i,j,0,t}\left(x\right) \leftarrow \hat{L}_{i,j,t}\left(x\right)\ \text{and} \ \hat{L}^\text{CRC}_{i,j,0,t}\left(x\right) \leftarrow \hat{L}^\text{CRC}_{i,j,t}\left(x\right). \nonumber
        \end{equation}
        \While{$u \leq T^\textnormal{I}$}{
            Compute $\hat{\hat{L}}^\text{CRC}_{i,j,t}\left(x\right)$ according to Eq. \eqref{received importance indication check bits}. \\ \label{start}
            \If{$\hat{\hat{L}}^\textnormal{CRC}_{i,j,t}\left(x\right) == \hat{L}^\textnormal{CRC}_{i,j,t}\left(x\right)$}{
                $break$
            }
            \Else{
                Request retransmission from transmitter to obtain $\hat{L}_{i,j,u,t}\left(x\right)$  and $\hat{L}^\text{CRC}_{i,j,u,t}\left(x\right)$. \\
            }
            Update $\hat{L}_{i,j,t}\left(x\right)$ and $\hat{L}^\text{CRC}_{i,j,t}\left(x\right)$ according to Eq. \eqref{majority voting} based on the majority voting. \\
            $u \leftarrow u+1$ \\ \label{end}
        }
        Initialize the retransmission counter $u \leftarrow 1$, and the packet data for the 0th retransmission: 
        \begin{equation}
            \hat{S}^{\text{P}}_{i,j,0,t}\left(x\right) \leftarrow \hat{S}^{\text{P}}_{i,j,t}\left(x\right)\ \text{and} \ \hat{S}^{\text{CRC}}_{i,j,0,t}\left(x\right) \leftarrow \hat{S}^{\text{CRC}}_{i,j,u,t}\left(x\right). \nonumber
        \end{equation}
        \While{$u \leq f_{\mathbf{l} \to T^\textnormal{R}}\left(\hat{\mathbf{l}}_{i,j,t}\right)$}{
            Similar to \ref{start}-\ref{end} in Algorithm \ref{received priority update}.
        }
    }
    \KwOutput{The received encrypted packet set $\left\{\mathbf{s}^\text{I}_{i,j,t}\right\}$.}

\end{algorithm}

Additionally, to fully utilize the information from retransmissions, the majority voting (MV) method is employed to combine the codewords from each retransmission \cite{MV}. Let $U$ represent the total retransmission times, and $c_{v,k}$ represent the $v$-th codeword of the $u$-th retransmission. Let $\hat{c}_v$ is the $v$-th codeword after voting, which can be expressed as
\begin{equation}
    \hat{c}_v=\underset{b}{\arg\!\min} \sum_{u=1}^{U}f_\mathbf{I}\left(c_{v,k}=b\right), \label{majority voting}
\end{equation}
where $b$ is a binary codeword, taking values of either 0 or 1. $f_\mathbf{I} \left(\cdot\right)$ represents the indicator function, which equals 1 when $c_{v,k}=b$ is true, and 0 otherwise.

\begin{figure*}[htbp]
    \centering
        \subfloat[Network structure of semantic transmitter]{\includegraphics[width=84mm]{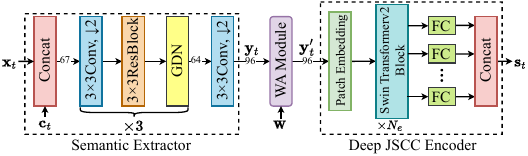}}%
    \hfill
        \subfloat[Network structure of semantic receiver]{\includegraphics[width=94mm]{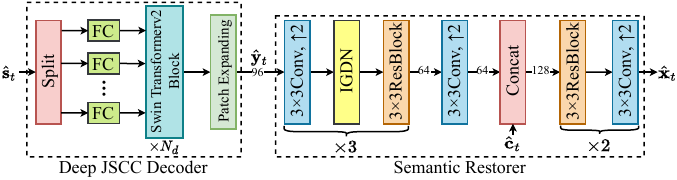}}
    \caption{Network structures of APVST. $k\times k$ Conv is a convolution with $k\times k$ filters, and the output channels of convolution are given on horizontal line. $\uparrow2$ and $\downarrow2$ indicate upsampling and downsampling with a stride of 2. GDN denotes the Generalised Divisive Normalization in \cite{density_modeling_images}, IGDN denotes the inverse operation of GDN.}
    \vspace{-2mm}
    \label{network of system}
\end{figure*}

\section{Enabled Modules and Internal Structure of APVST} \label{enabled modules and internal structure}

This section introduces the enabled modules and internal structure of the APVST network. Since the structures of primary and motion links are similar, the primary link will be introduced in detail.

\subsection{Enabled Modules for APVST}

In the description of the transmission process of the system model, we mention several key modules, including the WA module and latitude adaptive module, which are crucial for the efficient transmission of panoramic videos. In the following, a detailed explanation of the working principles of these modules and their roles within the system are provided.
\subsubsection{WA module}
For a more accurate assessment of panoramic video quality in observational spaces, the metrics WS-PSNR and WS-SSIM are utilized \cite{WS_PSNR}, \cite{WS_SSIM}. These metrics integrate the distortion interplay between projection planes and the viewer's FoV, mapping spherical distortions onto 2D plane distortions. This method more precisely reflects the immersive quality of the viewer's experience.


Inspired by the spatial attention module which is proposed in \cite{CBAM}, to obtain higher WS-PSNR and WS-SSIM which means the immersive experience of users, initial weight map $\mathbf{w}$ and feature map $\mathbf{y}_t$ are combined to achieve weighted spatial attention by the neural networks. The weighted spatial attention map $\mathbf{M}_t$ can be obtained as
\begin{align}
\mathbf{M}_t &\!=\!\sigma\!\left(\text{Conv}_{3\times3}^2\left\{\text{AvgPool}^4\left(\mathbf{w}\!\right);\text{MaxPool}^4\left(\mathbf{w}\right);\mathbf{y}_t\right\}\right) \nonumber \\
&\!=\!\sigma\left(\text{Conv}_{3\times3}^2\left\{\mathbf{F}_\text{Avg};\mathbf{F}_\text{Max};\mathbf{y}_t\right\}\right),\label{eq8}
\end{align}
where $\text{AvgPool}^4$ and $\text{MaxPool}^4$ denote that the initial weight maps $\mathbf{w}=\{w\left(p,q\right)\}\in\mathbb{R}^{H \times W}$ is downsampled by successive average pooling and maximum pooling four times, respectively. $H$ and $W$ represent the height and width of panoramic frames, respectively. Besides, a dimension is inserted into the pooling results to represent the number of channels, so the generated feature weight distributions are $\mathbf{F}_\text{Avg}\in\mathbb{R}^{1\times\frac{H}{16}\times\frac{W}{16}}$ and $\mathbf{F}_\text{Max}\in\mathbb{R}^{1\times\frac{H}{16}\times\frac{W}{16}}$, respectively. $\text{Conv}_{3\times3}^2$ denotes two successive convolution operations with convolution kernel of $3\times3$, and $\sigma$ is the Sigmoid activation function. Besides, the weight values $w\left(p,q\right)$ decrease gradually from the equator to the poles, and the distribution of $w\left(p,q\right)$ is given as 
\begin{equation}
    w\left(p,q\right)=\cos{\left(\left(p-\frac{H}{2}+\frac{1}{2}\right)\times\frac{\pi}{H}\right)}.\label{eq9}
\end{equation}

The feature information after deflation is obtained by multiplying the feature map with the points corresponding to the weighted spatial attention map. This process can be described as $\mathbf{y}_t^\prime=\mathbf{y}_t\otimes\mathbf{M}_t$.

Leveraging the powerful learning capacities of neural networks, the weight maps are fused into the feature maps. This approach mirrors the attention mechanism by allocating weights to specific points on the feature map, thereby intensifying focus on areas with higher weights during the information restoration phase. As a result, this method significantly enhances the quality of the viewer's immersive experience as measured by WS-PSNR and WS-SSIM.

\subsubsection{Latitude Adaptive Module}
Due to the characteristics of ERP, it stretches the pixels at different latitudes to different degrees. According to \cite{End_to_End_Optimized_360_Image_Compression}, the higher latitude needs more pixels and, consequently, it needs more bits to represent high-latitude pixels if a uniform compression strategy is used. 

To reduce the information redundancy of transmission and improve the transmission efficiency of panoramic video, we propose the latitude adaptive network. For such a network, the adaptive weight feature map $\boldsymbol{\omega}_t=\left\{\omega_t(m,n)\right\}\in\mathbb{R}^{\frac{H}{16}\times\frac{W}{16}}$ is designed, which can be expressed as
\begin{equation}
\omega_t(m,n)=\eta_t(m,n){w}_\text{AAP}(m,n)+1-\eta_t(m,n),\label{eq16}
\end{equation}
where $\omega_t(m,n)$ denotes the adaptive weight feature value corresponding to the feature point $\left(m,n\right)$ at time $t$. ${w}_\text{AAP}(m,n)$ denotes the weight value of $\mathbf{w}$ after the adaptive average pooling (AAP) operation. $\eta_t(m,n)$ denotes the adaptive weight factor value of weight factor map $\boldsymbol{\eta}_t=\left\{\eta_t(m,n)\right\}\in\mathbb{R}^{\frac{H}{16}\times\frac{W}{16}}$, which is learned by the network according to the magnitude of entropy information. It can be expressed as $\boldsymbol{\eta}_t=f_\text{AW}\left(\mathbf{e}_t\right)$, where $f_\text{AW}\left(\cdot\right)$ denotes the function of adaptive weight factor network. 

In order to achieve the adaption of feature information dimension in latitude, the dimensions of the information in the wireless channel cannot exceed the limit of the maximum dimension given in the corresponding latitude. This constraint can be expressed as
\begin{equation}
l_t(m,n)\le \omega_t(m,n)\times {\max}\left(\mathcal{Q}\right),\label{eq17}
\end{equation}
where $l_t(m,n)$ denotes the information dimension corresponding to feature point $\left(m,n\right)$ by quantizing, which can be expressed as the number of channels. ${\max}\left(\mathcal{Q}\right)$ is the maximum value of the quantized set, which is given in Section \ref{section experiment}.
Therefore, the loss function of this module is expressed as
\begin{equation}
\mathcal{L}_t^\text{la}=\sum_{m}\sum_{n}{{\max}\left(0,\omega_t(m,n)\times {\max}\left(\mathcal{Q}\right)-l_t(m,n)\right)}.\label{eq18} 
\end{equation}

However, the quantization operation tends to cause gradient dispersion \cite{End_to_end_optimized_image_compression}, we convert the dimensional restriction to the entropy restriction in the training phase, i.e,
\begin{equation}
    \mathcal{L}^\prime{_t^\text{la}}=\sum_{m}\sum_{n}{{\max}\left(0,\omega_t(m,n)\times{\max}\left(\mathbf{e}_t\right)-e_t(m,n)\right)}.\label{loss of la} 
\end{equation}

The latitude adaptive module allows the entropy model to calculate the entropy of feature maps across various latitudes more precisely, facilitating the efficient transmission of panoramic videos. Additionally, by implementing an appropriate loss function, the APVST network optimizes during training to minimize channel bandwidth usage while maintaining the quality of transmission.

\subsection{The Internal Structure of APSVT}
\begin{figure}[htbp]
    \centering
    \vspace{-2mm}
    \subfloat[]{\includegraphics[width=35mm]{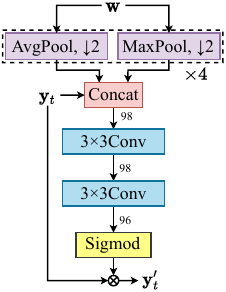} \label{network of wa}} 
    \quad
    \quad 
    \quad
    \subfloat[]{\includegraphics[width=23mm]{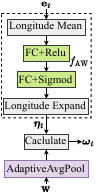} \label{network of la}} 
    \caption{Detailed structures of (a) WA Module, and (b) Latitude Adaptive Module.}
    \label{network of other modules}
\end{figure}

The disassembled parts of the internal structure are shown in Fig.~\ref{network of system} and Fig.~\ref{network of other modules}.
\subsubsection{The overall structure of APSVT}

According to \cite{DCVC}, the current panoramic frame $\mathbf{x}_t$ is firstly concatenated with the context $\mathbf{c}_t$. Semantic extraction is performed using a series of downsampling convolutions, a residual structure (Resblock), and GDN. In contrast, during semantic recovery, the feature map undergoes successive upsampling through convolutions with IGDN and Resblocks. Ultimately, ${\hat{\mathbf{c}}}_t$ is integrated with the upsampling result, and the number of channels is adjusted using a convolution layer and Resblocks. The Deep JSCC architecture incorporates symmetric encoding and decoding strategies, and power normalization is performed before the information is transmitted into the channel. To enhance its ability to capture long-term correlations, the Deep JSCC utilizes the swin transformer v2 as the main framework for both its encoder and decoder, with multi-head self-attention (MHSA) as the network's backbone. Differing from the original swin transformer, swin transformer v2 adopts cosine similarity and a nonlinear relative position bias, improving the network's resilience across various downstream tasks \cite{swin_transformerv2}.

The Deep JSCC encoder is guided by the entropy model and the latitude adaptive module to achieve the variable-length encoding of the feature map. The information dimension $l_t(mn)$ is derived by quantizing the entropy $e_t(m,n)$ into the set of quantized values, which can be expressed as $l_t(m,n)\!\in\!\mathcal{Q}\!=\!\left\{q_1,q_2,\cdots,q_I\right\}$ with size $I$. By multiple learnable fully connected (FC) layers, the feature point dimension is adjusted to $l_t(m,n)$ to achieve the variable-length encoding.

\subsubsection{Weight attention module}
The proposed WA module deflates the feature maps appropriately to enhance the final video quality evaluation metrics, i.e. WS-PSNR and WS-SSIM, which structure is illustrated in Fig.~\ref{network of wa}. To maximize the retention of weight map information, the weight map $\mathbf{w}$ and the feature map $\mathbf{y}_t$ are concatenated. Due to the different sizes of $\mathbf{w}$ and $\mathbf{y}_t$, two different pooling downsampling methods (four consecutive average pooling and maximum pooling) are employed to act on $\mathbf{w}$. The obtained result is concatenated with $\mathbf{y}_t$, and the weighted spatial attention map $\mathbf{M}_t$ is finally obtained by the Sigmoid activation function after two layers of convolution operation. The first convolution layer is designed to learn the relationship between $\mathbf{M}_t$ and $\mathbf{y}_t$, while the second layer adjusts the number of channels. The Sigmoid activation function is used to restrict the weight values between 0 and 1.

\subsubsection{Latitude adaptive module}
The application of the latitude adaptive module reduces information redundancy and enhances the transmission efficiency of panoramic videos. By utilizing function $f_{\text{AW}}\left(\cdot\right)$, the adaptive weight feature map $\boldsymbol{\omega}_t$ is derived from the entropy map $\mathbf{e}_t$ and the weight map $\mathbf{w}$. The detailed module structure is shown in Fig.~\ref{network of la}. Given that the feature map is more latitude-sensitive, a conversion from $\mathbf{e}_t\in\mathbb{R}^{\frac{H}{\mathbf{16}}\times\frac{W}{\mathbf{16}}}$ to $\mathbb{R}^{\frac{H}{\mathbf{16}}}$ is performed by averaging over longitude. Subsequently, the weight factor map $\boldsymbol{\eta_t}$ is generated through FC layers using ReLU and Sigmoid activation functions and an expansion along the longitude is applied to transform $\mathbb{R}^{\frac{H}{\mathbf{16}}}$ back to $\mathbb{R}^{\frac{H}{\mathbf{16}}\times\frac{W}{\mathbf{16}}}$. Finally, $\boldsymbol{\eta_t}$ and  $\mathbf{w}$ are merged to compute $\boldsymbol{\omega}_t$ according to Eq.~\eqref{eq16}.

\section{Experimental Results and Analysis} \label{section experiment}
This section presents the training and testing datasets, details of parameter settings, and analysis of experimental results. By simulation analysis, we demonstrate the superiority of semantic communication in cross-layer transmission compared with other transmission schemes.

\subsection{Experimental Setup}
\subsubsection{Parameters Settings of APVST}
The proposed APVST network is trained on the panoramic video dataset of 360VDS \cite{omnidirectional_video_super_resolution}, the dataset of which contains 590 panoramic videos in ERP. The panoramic frames are resized to $512\times256$ pixels and randomly flipped during training. We evaluate the APVST on the VR scene dataset \cite{gaze_prediction_immersive_videos}, which contains 208 panoramic videos. For the first frame (I-frame) coding of a GoP, we apply nonlinear transform source-channel coding (NTSCC) \cite{ntscc}.

In the APVST experiment, the number of blocks in swin transformerv2 is set to $N_e=N_d=4$. 8 heads and $8\times8$ window size are used in MHSA. The quantization set in Deep JSCC is set as $\mathcal{Q}=\{0,2,4,6,8,10,16,20,26,32,20,$ $48,56,64,80,96\}$. In the model training and testing processes, we set the GoP sizes as $N=7$ and $N=4$, respectively. Adam optimizer is adopted as the optimizer. Other parameters are set as the learning rate is ${10}^{-4}$, the training batch size is 8, and the testing batch size is 1. The whole APVST model was trained on a single A40 GPU for 4 days.

\subsubsection{Parameters Settings of Cross-Layer Design}
During cross-layer transmission, the semantic importance is divided into 5 levels for verification. The height and width of panoramic videos are set to $H\times W = 512 \times 1024$ during testing. The granularity of grouping and packet size is set to $m_H = m_W = 8$ and $1024$, respectively. The number of subcarriers $K$ in OFDMA is set to $256$. The transmission band is set within a frequency band ranging around a center frequency of $f_c=2.6\text{GHz}$ with a bandwidth of $B=200\text{MHz}$. The transmit power spectral density and noise power spectral density are set to $-53 \text{dBm/Hz}$ and $-143 \text{dBm/Hz}$, respectively. The settings for different levels of importance indication are shown in Table \ref{tab:para_settings}.

\begin{table}[h]
    \renewcommand\arraystretch{1.3}
    \centering
    \caption{Parameters Settings of Different Importance Indication}
    \label{tab:para_settings}
    \begin{tabular}{|c|c|c|c|c|c|} 
        \hline
            \textbf{Level of importance indication} & 1 & 2 & 3 & 4 & 5 \\ 
        \hline
            \textbf{CRC bits length} & 4 & 8 & 16 & 24 & 32 \\ 
        \hline
            \textbf{LDPC coding efficiency} & 2/3 & 2/3 & 1/2 & 1/2 & 1/3 \\ 
        \hline
            \textbf{Maximum times of retransmission} & 2 & 3 & 4 & 6 & 10 \\ 
        \hline
    \end{tabular}
\end{table}

\subsubsection{Comparison Schemes} 
We compare the performance of cross-layer transmission for panoramic videos by semantic communication schemes and traditional communication schemes. For the semantic communication-based cross-layer transmission schemes, we consider ``APVST'', ``APSVT  (w/o WA)'', and ``DVST'' \cite{DVST} in the semantic representation layer of CLESC, where ``APSVT  (w/o WA)'' represents the APVST network without WA module. For traditional communication-based cross-layer transmission schemes, we consider ``H.264'' \cite{H_264} and ``H.265'' \cite{H_265} in the application layer, both of these are efficient video source encoding schemes. All schemes employ LDPC for channel coding.

\subsubsection{Evaluation Metrics}
For panoramic videos, the distinctive properties of ERP lead to different distortions in various regions. To address this, we have chosen pixel-level metric WS-PSNR and perceptual metric WS-SSIM for evaluation, effectively reflecting the regional visual distortions inherent to panoramic videos. WS-PSNR focuses on global average distortion, whereas WS-SSIM offers a more detailed evaluation of perceptual distortion. Moreover, we have integrated the deep learning-based perceptual metric, learned perceptual image patch similarity (LPIPS), which simulates the human perception process. LPIPS outputs a loss score ranging from 0 to 1, where lower scores indicate reduced visual distortion, thus providing an effective measure of visual fidelity. This comprehensive use of metrics not only enhances the accuracy of our evaluations but also aids in thoroughly assessing the performance of semantic communication in cross-layer transmission.

\begin{figure*}[tbp]
    \centering
        \subfloat[]{\includegraphics[width=60mm]{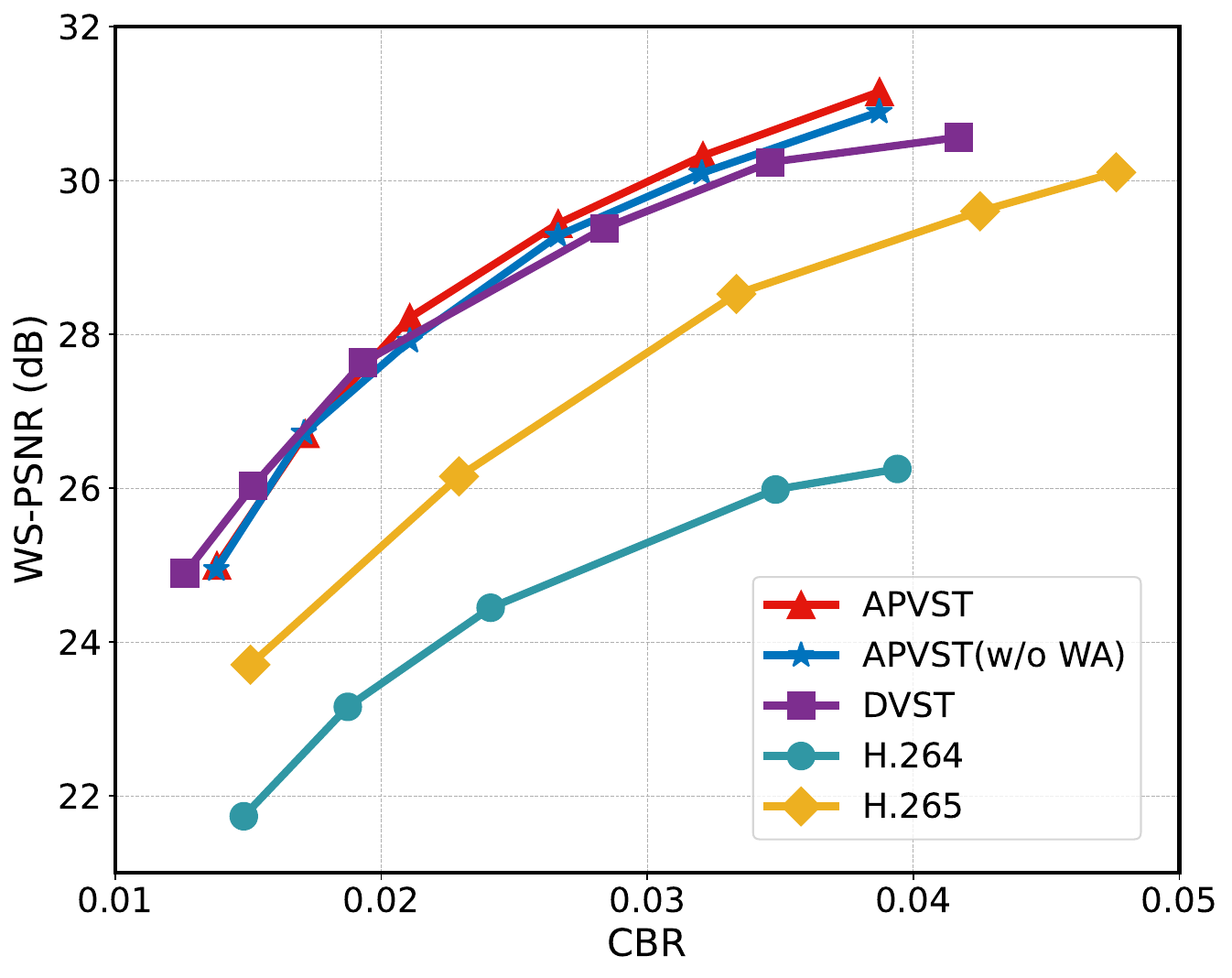} \label{semantic wspnsr vs cbr}}
    \hfill
        \subfloat[]{\includegraphics[width=58mm]{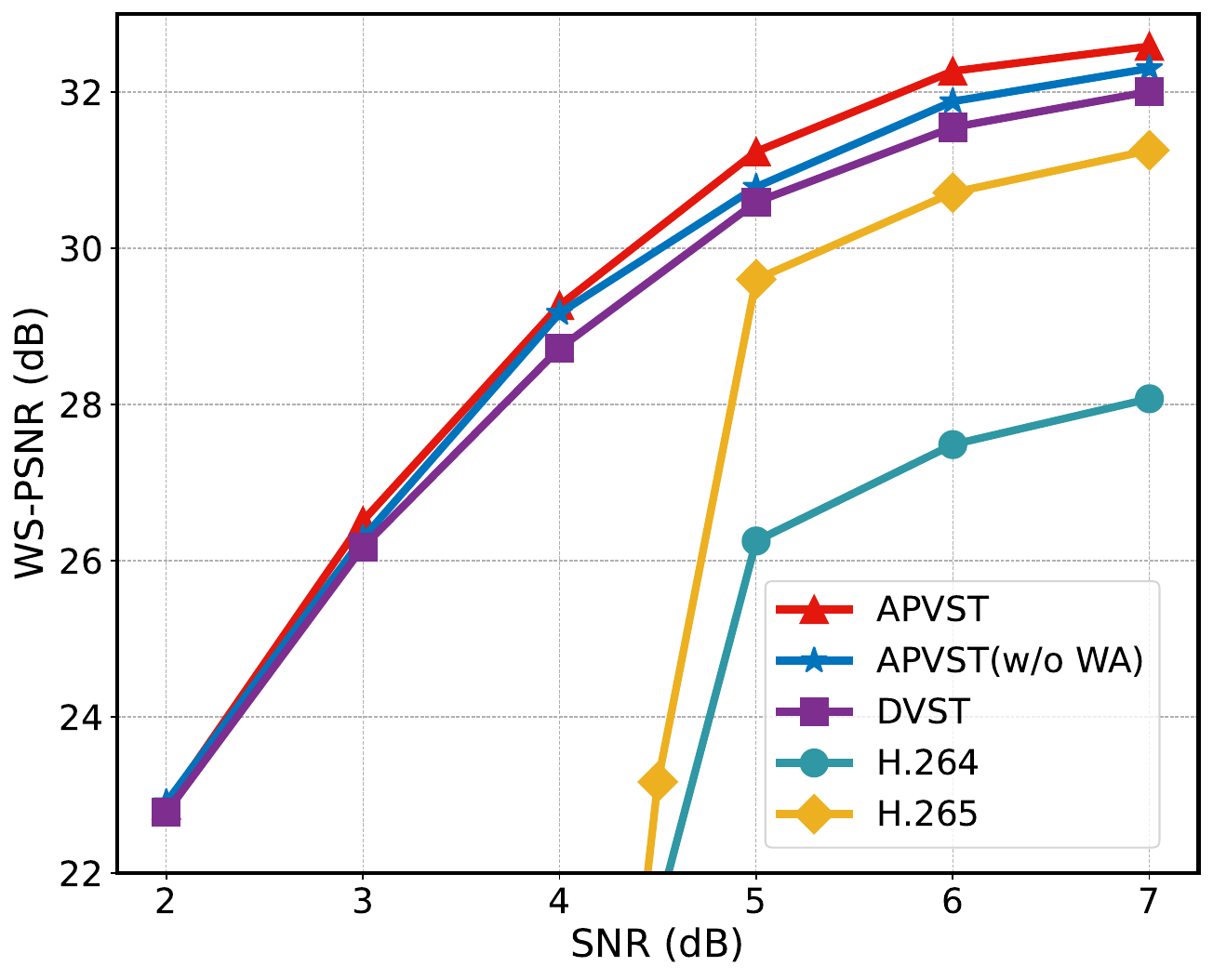} \label{semantic wspnsr vs snr}}
    \hfill
        \subfloat[]{\includegraphics[width=58mm]{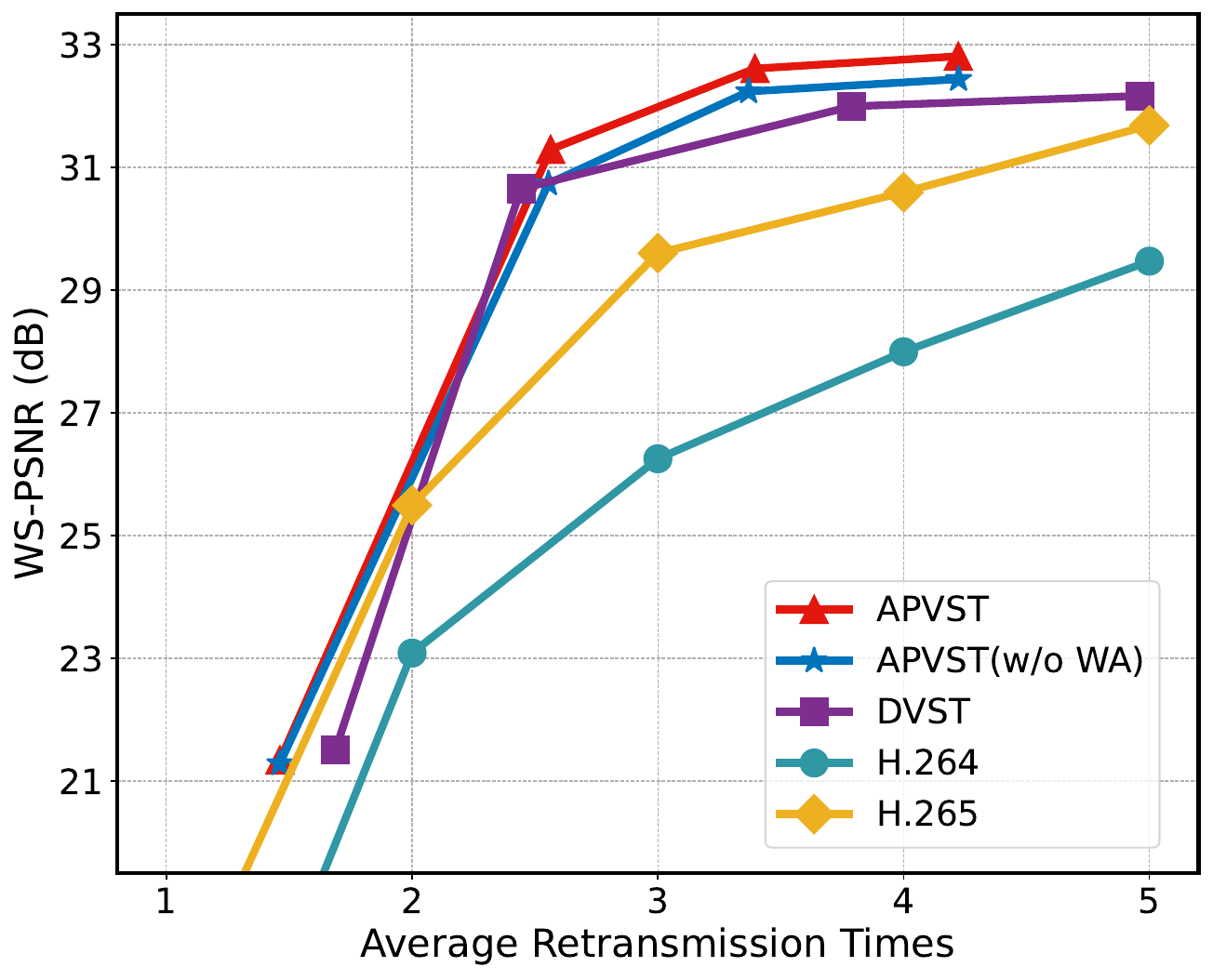} \label{semantic wspnsr vs re}}
    \caption{WS-PSNR performance vs. (a) CBR at SNR=5dB, (b) SNR at CBR=0.04, and (c) average retransmission times at SNR=5dB and CBR=0.04 with BPSK modulation.}
    \label{wspnsr vs others}
    \vspace{-2mm}
\end{figure*}

\begin{figure*}[tbp]
    \centering
        \subfloat[]{\includegraphics[width=60mm]{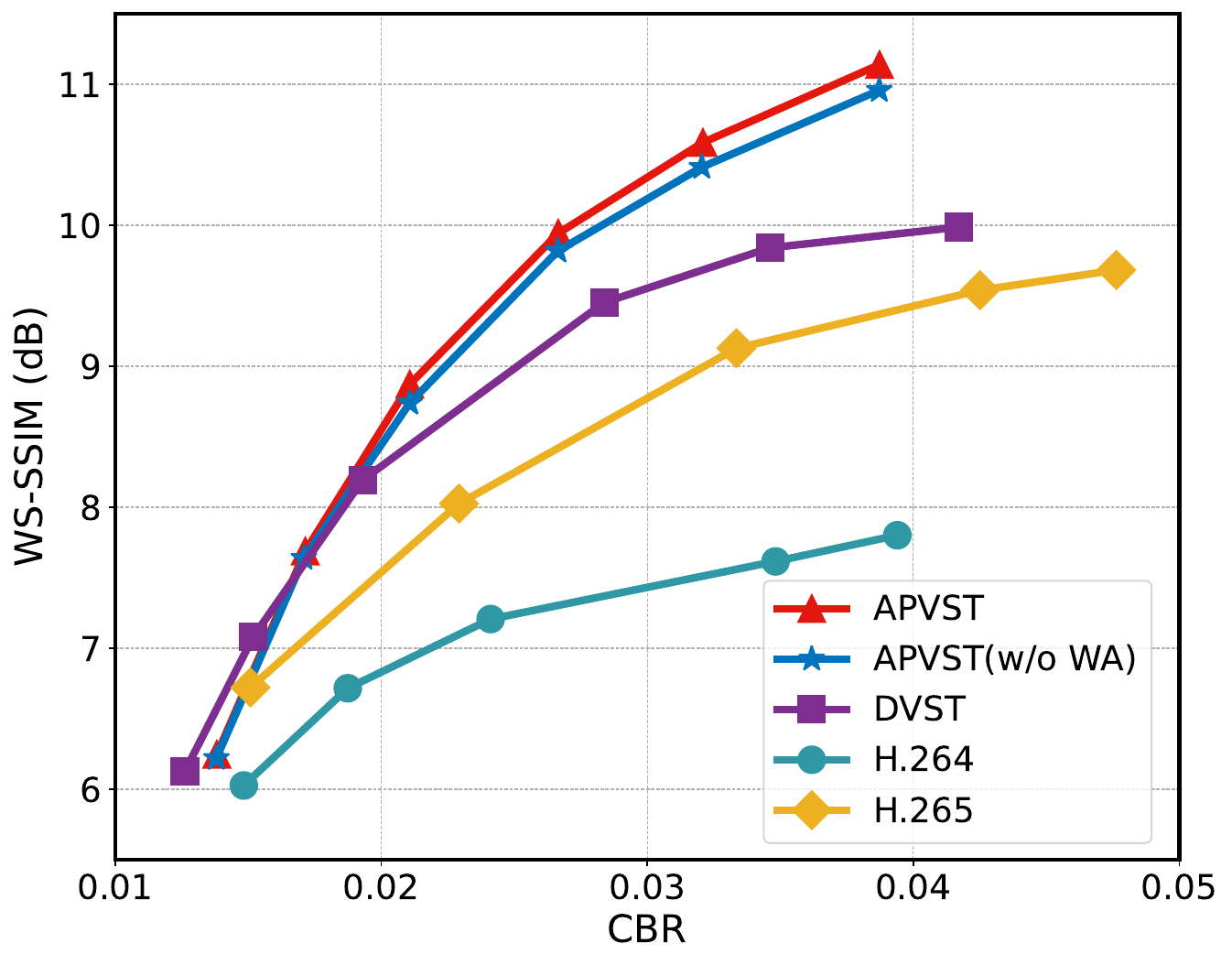} \label{semantic wsssim vs cbr}}%
    \hfill
        \subfloat[]{\includegraphics[width=58mm]{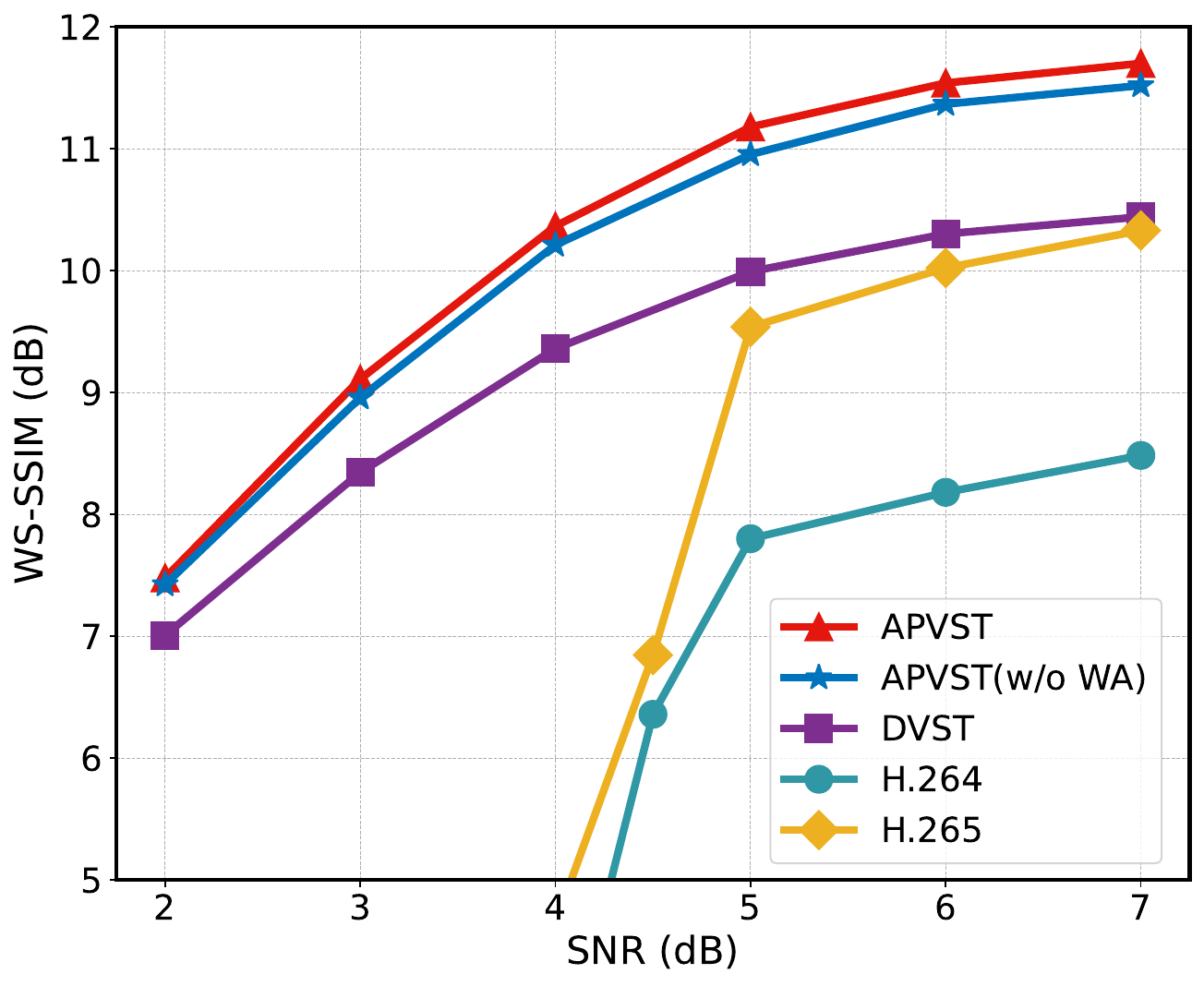} \label{semantic wsssim vs snr}}
    \hfill
        \subfloat[]{\includegraphics[width=58mm]{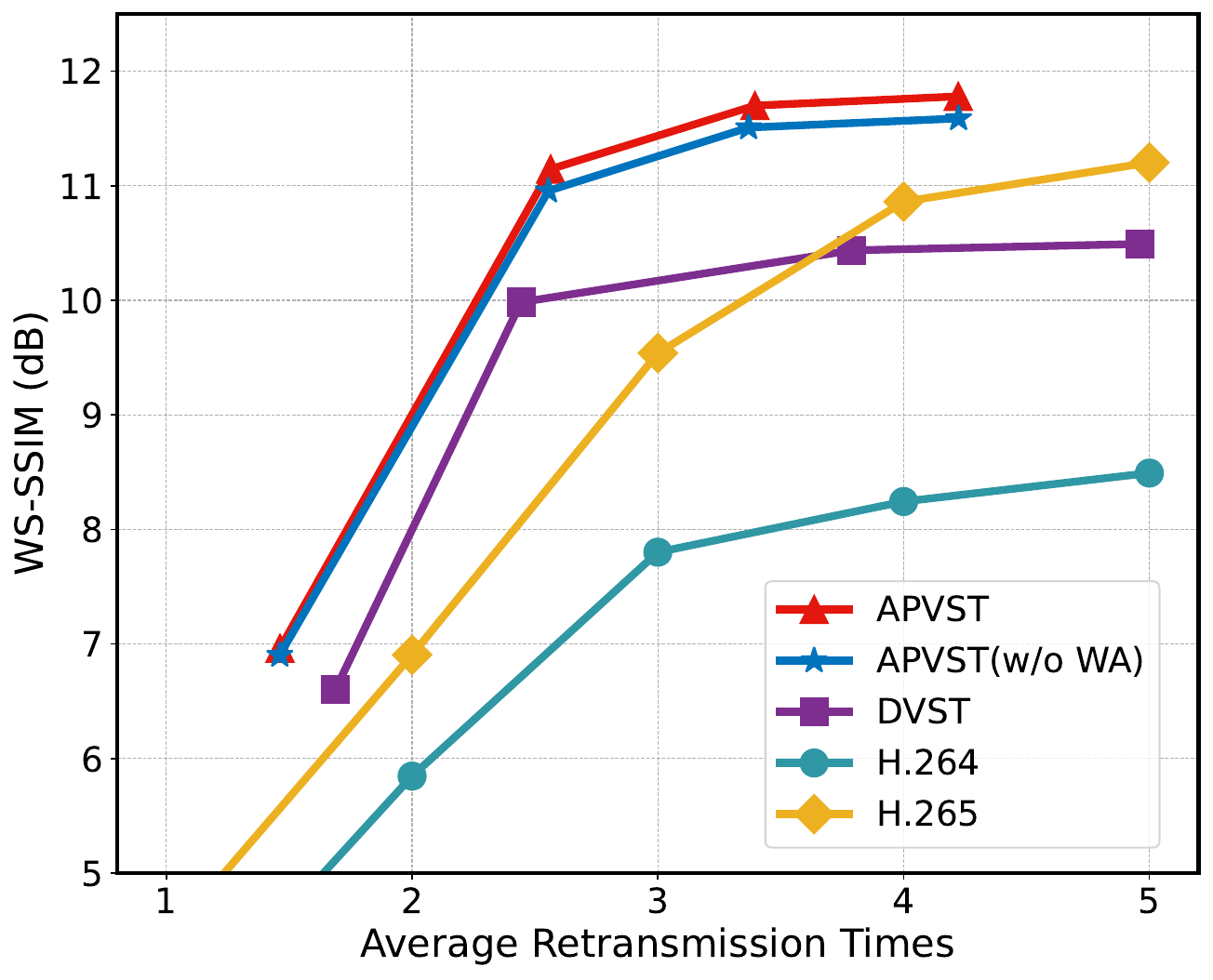} \label{semantic wsssim vs re}}
    \caption{WS-SSIM performance vs. (a) CBR at SNR=5dB, (b) SNR at CBR=0.04, and (c) average retransmission times at SNR=5dB and CBR=0.04 with BPSK modulation.}
    \label{wsssim vs others}
    \vspace{-2mm}
\end{figure*}

\vspace{-2mm}
\subsection{Experimental Analysis}
\subsubsection{WS-PSNR Performance}
Fig.~\ref{wspnsr vs others} shows the WS-PSNR achieved by APVST, APVST (w/o WA), DVST, H.264, and H.265 cross-layer transmission schemes versus CBR, SNR, and average retransmission times. For all schemes, binary phase shift keying (BPSK) modulation is adopted. By adjusting the end-to-end distance, different SNRs will be obtained. In this case, the reconstruction loss chosen for training the semantic network is WMSE. In traditional transmission schemes, we used a 1/2 rate (1024/2048) for LDPC coding and set the maximum retransmission number to 4. Experimental results show that semantic transmission schemes exhibit superior performance in cross-layer transmission compared to traditional schemes. This also means that semantic communication can be compatible with traditional communication frameworks, significantly enhancing system performance.

Fig.~\ref{semantic wspnsr vs cbr} and Fig.~\ref{semantic wspnsr vs snr} show that under all CBRs and SNRs, semantic cross-layer transmission schemes outperform traditional ones. In achieving the same WS-PSNR, the proposed APVST can reduce bandwidth consumption by 85\% and 33\% compared to H.264 and H.265, respectively. Moreover, as SNR decreases, semantic schemes effectively prevent the ``cliff effect'' that occurs in traditional schemes, ensuring a quality immersive experience at low SNR. However, it is observed that at low SNR, APVST and DVST achieve the same WS-PSNR. This is attributed to the WA module, which primarily learns from the semantic feature map $\mathbf{y}_t$ and the weight map $\mathbf{w}$ during the training phase, without incorporating noise-related information, thereby reducing the network's noise immunity.

\begin{figure*}[tbp]
    \centering
        \subfloat[]{\includegraphics[width=58mm]{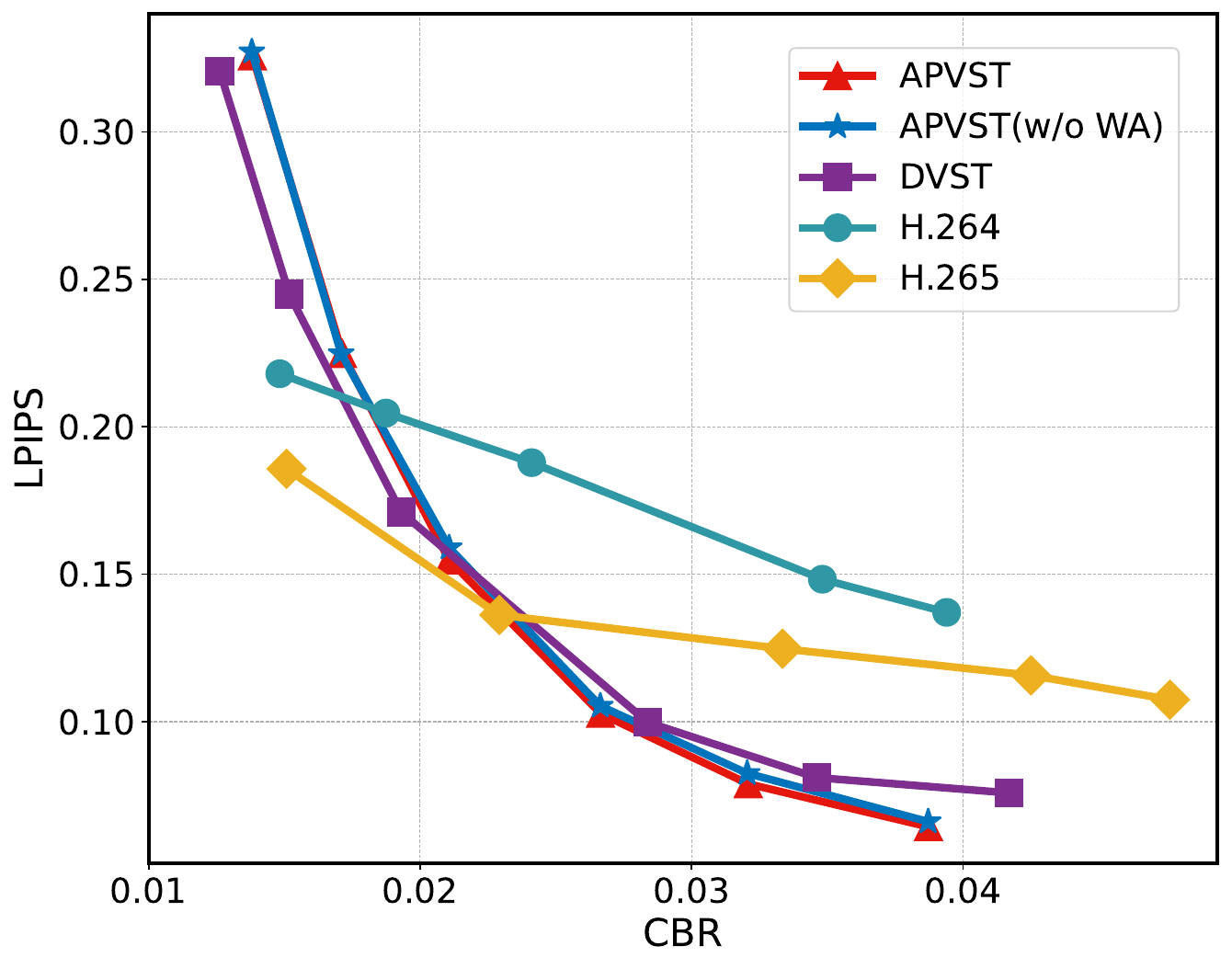} \label{semantic lpips vs cbr}}%
    \hfill
        \subfloat[]{\includegraphics[width=58mm]{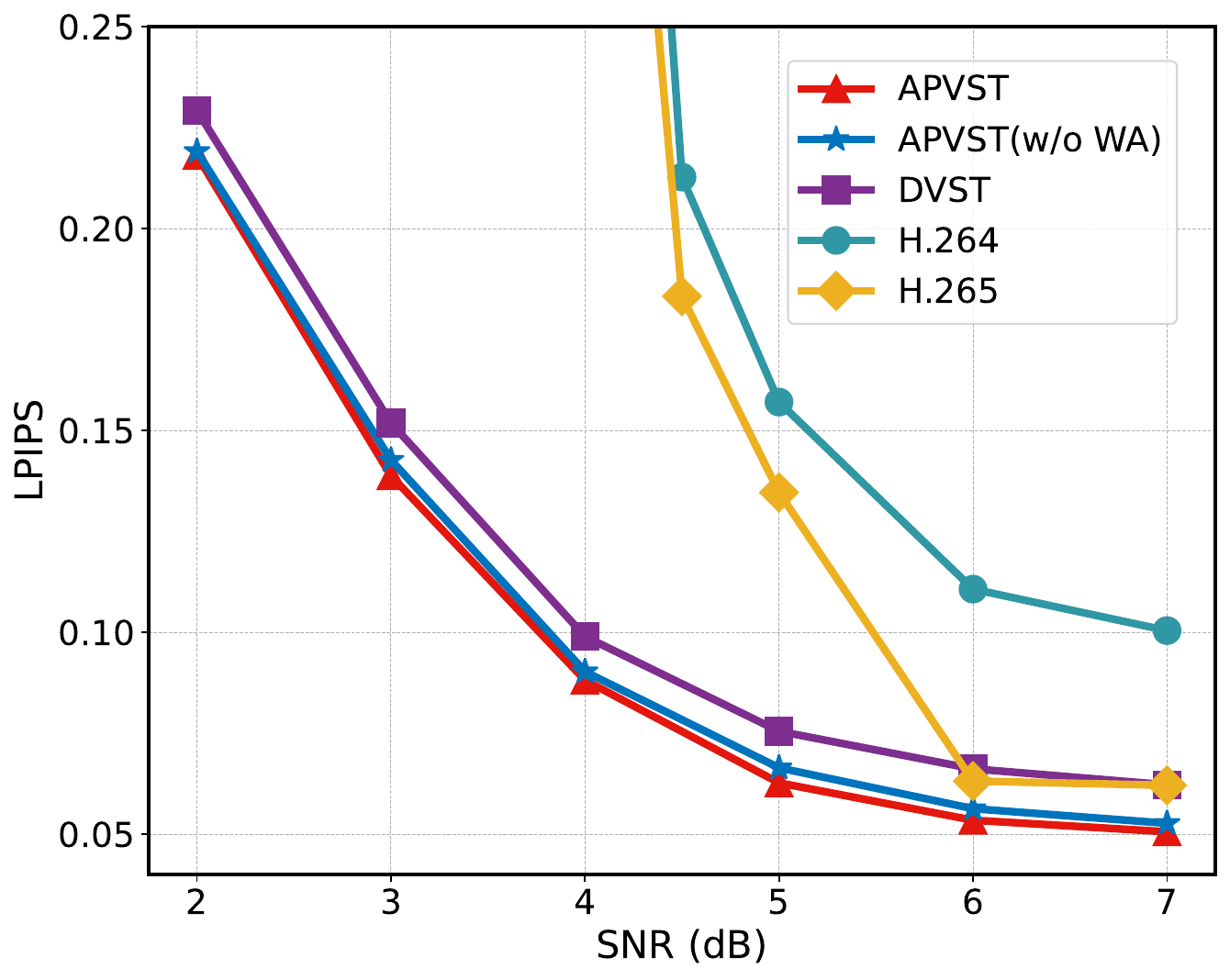} \label{semantic lpips vs snr}}
    \hfill
        \subfloat[]{\includegraphics[width=58mm]{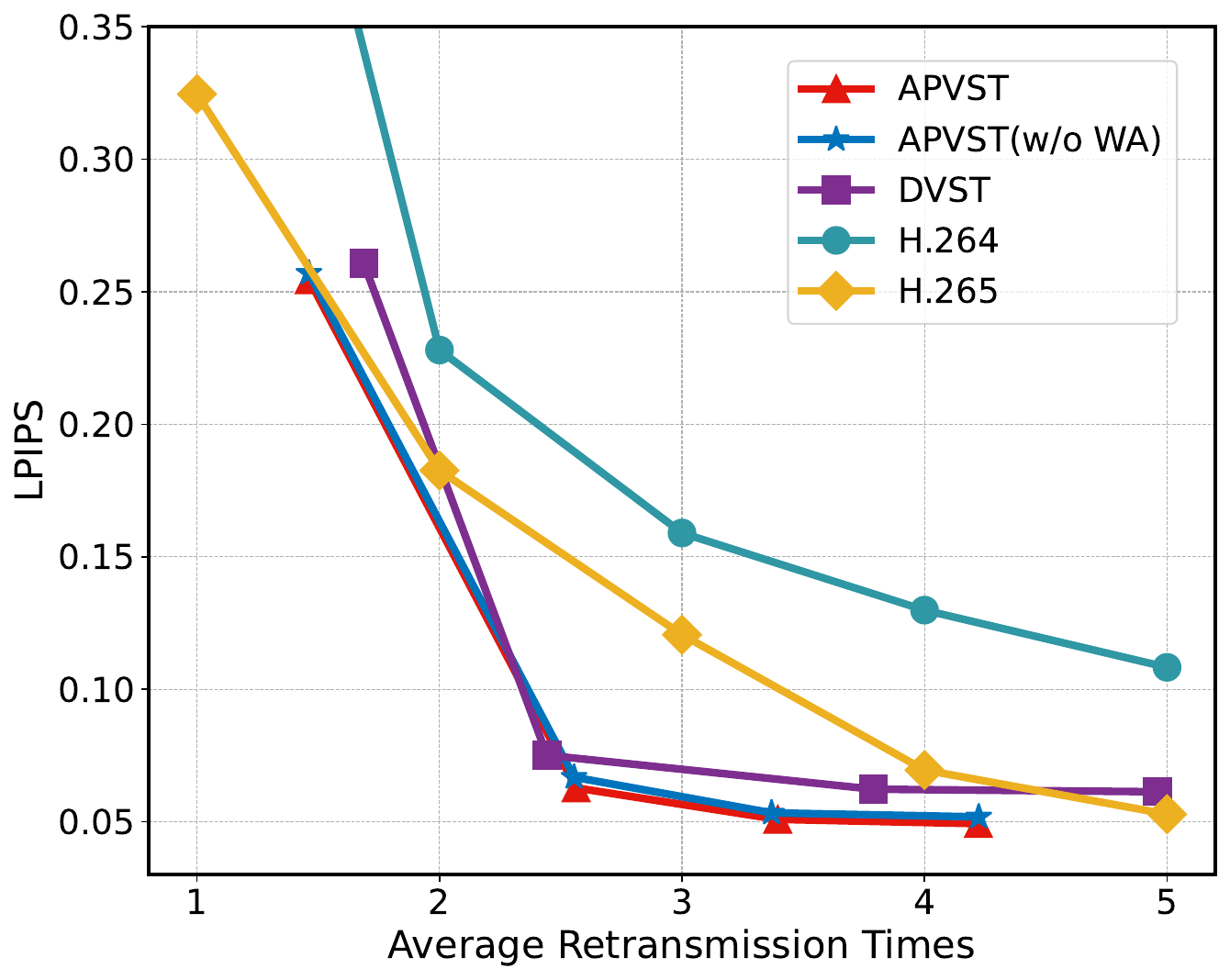} \label{semantic lpips vs re}}
    \caption{LPIPS performance vs. ((a) CBR at SNR=5dB, (b) SNR at CBR=0.04, and (c) average retransmission times at SNR=5dB and CBR=0.04 with BPSK modulation.}
    \label{lpips vs others}
\end{figure*}

Fig.~\ref{semantic wspnsr vs re} shows that the maximum retransmission count chosen at the data link layer also has an impact on transmission performance. Semantic transmission schemes can achieve near-optimal performance with an average retransmission time of about 3, increasing retransmission times beyond this point does not significantly benefit WS-PSNR but increases transmission latency instead. Traditional H.265 coding schemes require an average retransmission count of about 5 to achieve performance comparable to APVST. This further demonstrates that under the premise of achieving similar performance, semantic cross-layer transmission schemes can further reduce transmission latency and resource consumption, and enhance the efficiency of communication systems.

\begin{figure*}[ht]
    \begin{center}
    \begin{tabularx}{\textwidth}{m{1.2cm}<{\centering} m{3.5cm}<{\centering} m{3.5cm}<{\centering} m{3.5cm}<{\centering} m{3.8cm}<{\centering}}
         & Original & APVST & DVST & Semantic Importance Levels \\
        frame 1 & \includegraphics[width=0.21\textwidth]{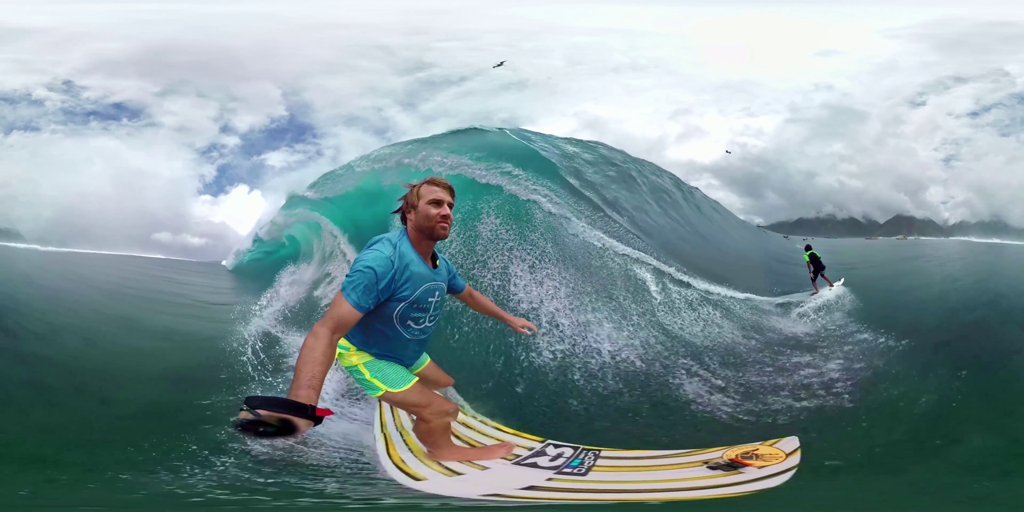} & 
        \includegraphics[width=0.21\textwidth]{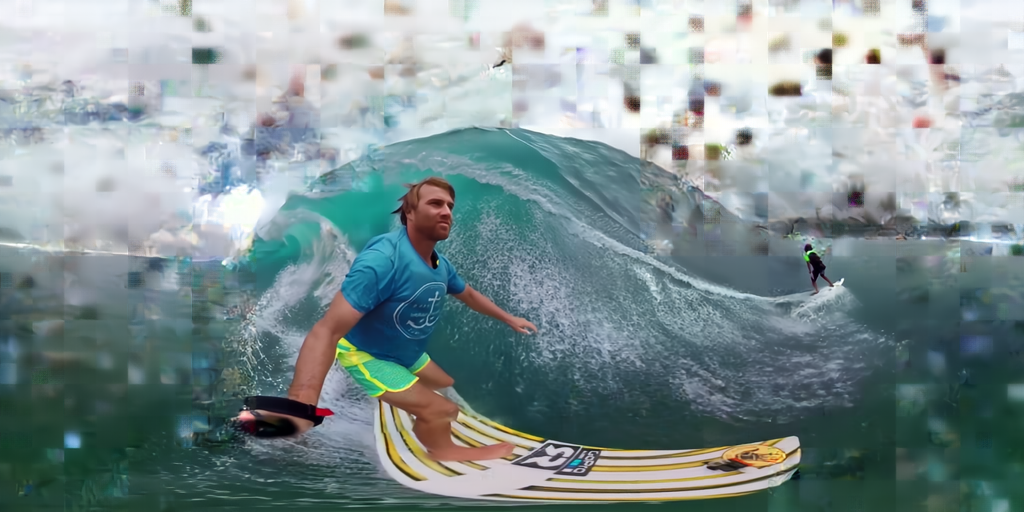} & 
        \includegraphics[width=0.21\textwidth]{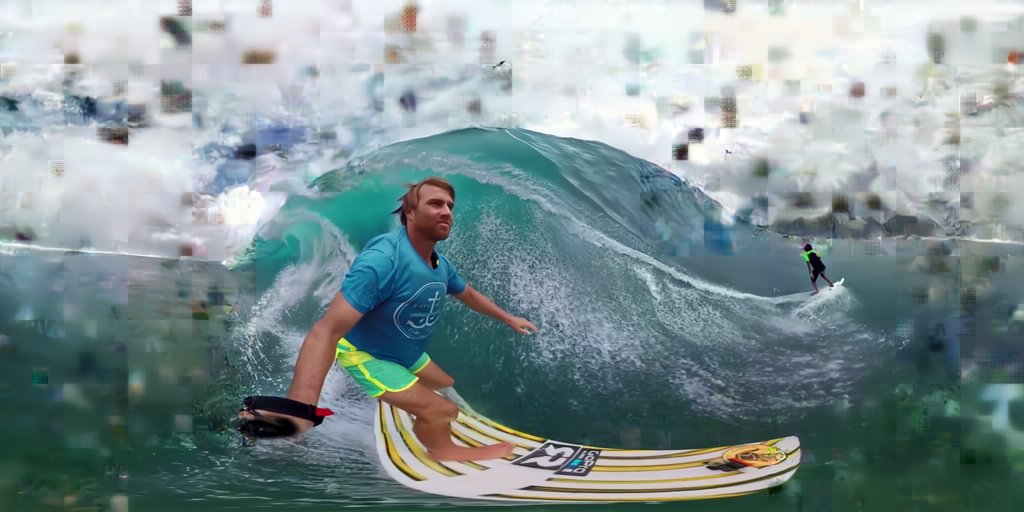} & 
        \includegraphics[width=0.23\textwidth]{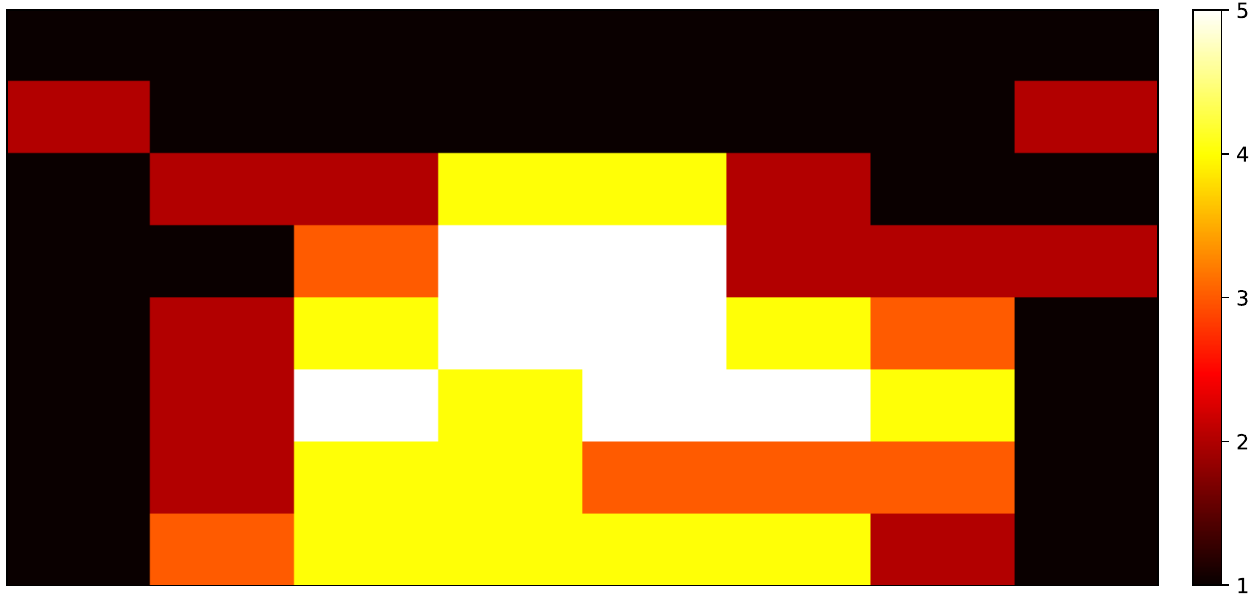}\\
        & CBR / WS-PSNR (dB) & 0.0962 / 22.184 & 0.0962 / 21.981 & \\
        frame 2 & \includegraphics[width=0.21\textwidth]{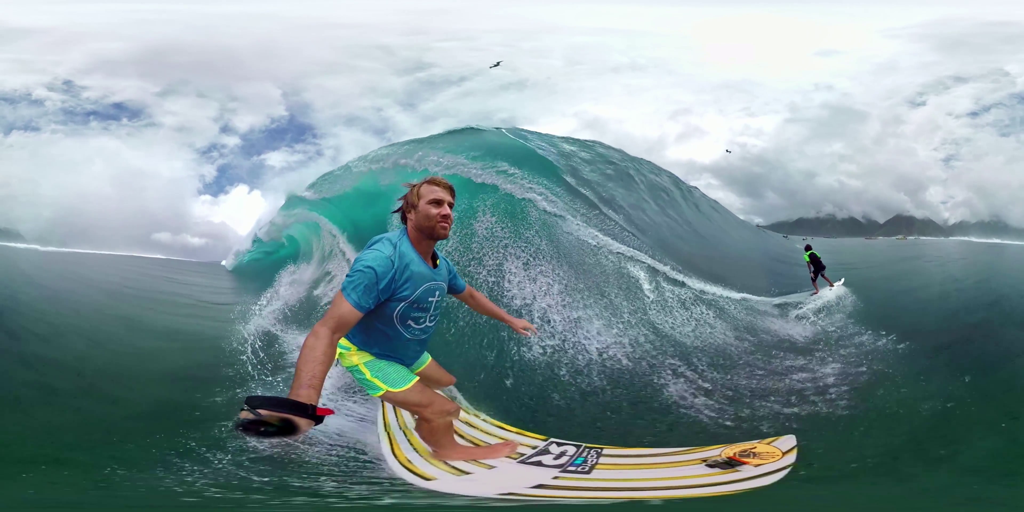} & 
        \includegraphics[width=0.21\textwidth]{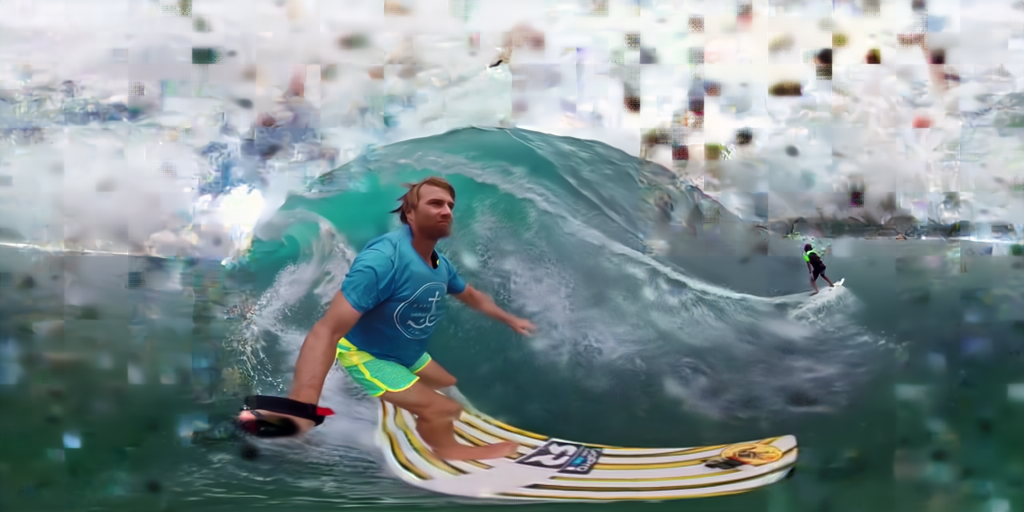} & 
        \includegraphics[width=0.21\textwidth]{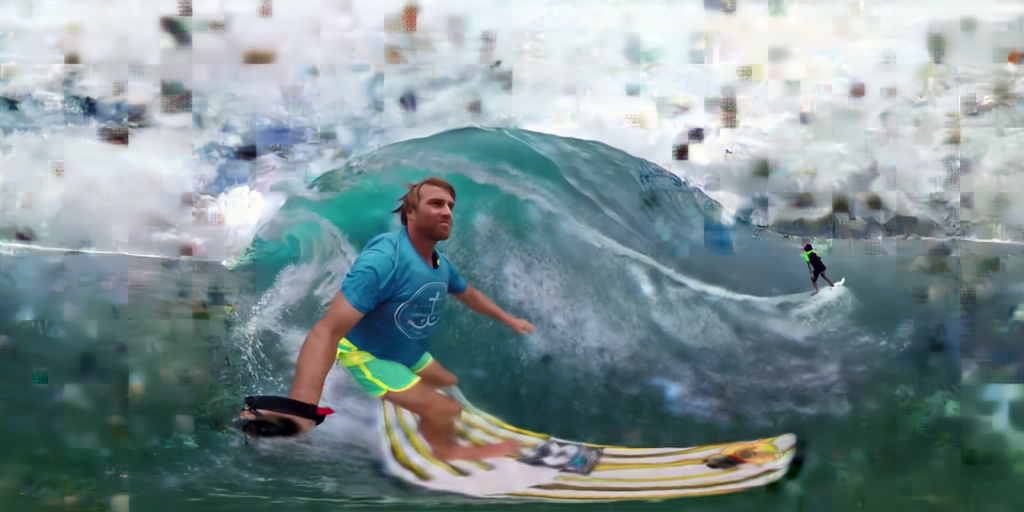} & 
        \includegraphics[width=0.23\textwidth]{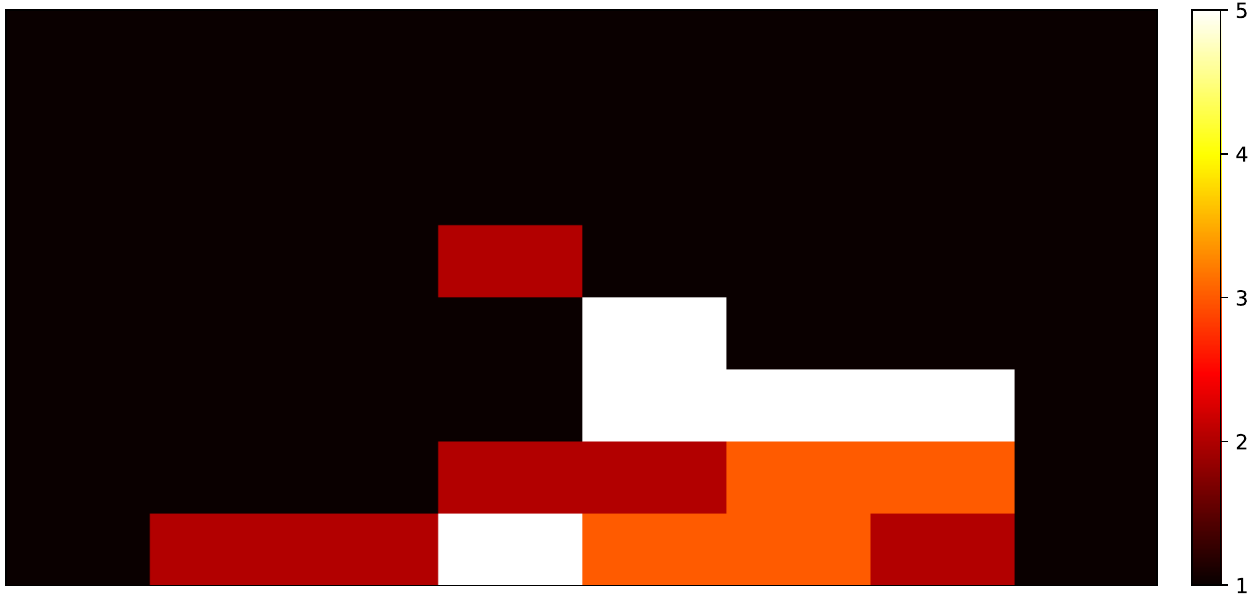}\\
        & CBR / WS-PSNR (dB) & 0.0124 / 21.358 & 0.0198 / 21.104 & \\
        frame 3 & \includegraphics[width=0.21\textwidth]{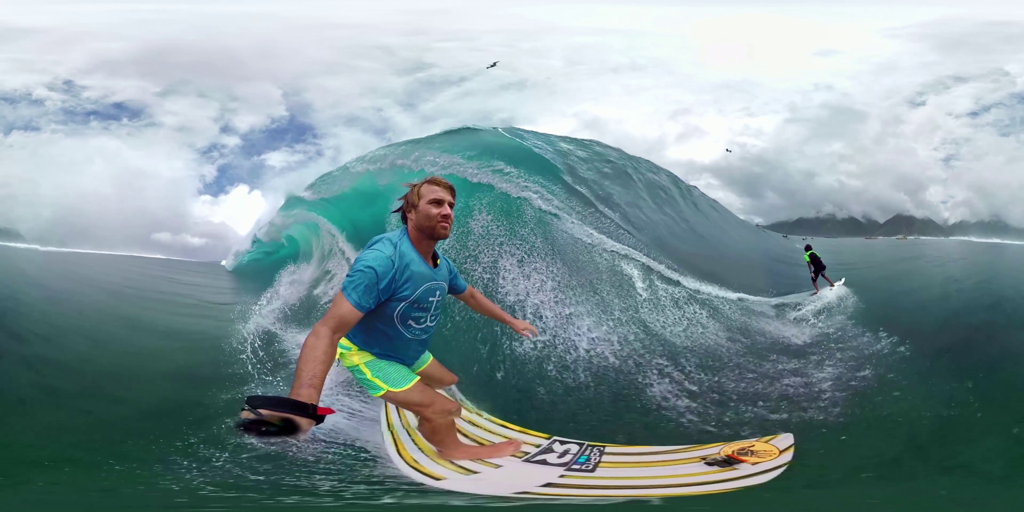} & 
        \includegraphics[width=0.21\textwidth]{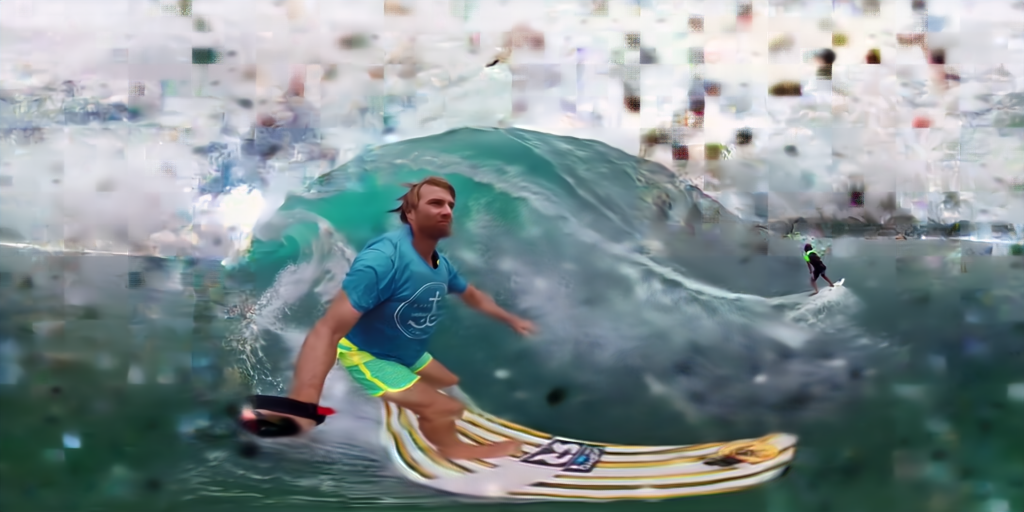} & 
        \includegraphics[width=0.21\textwidth]{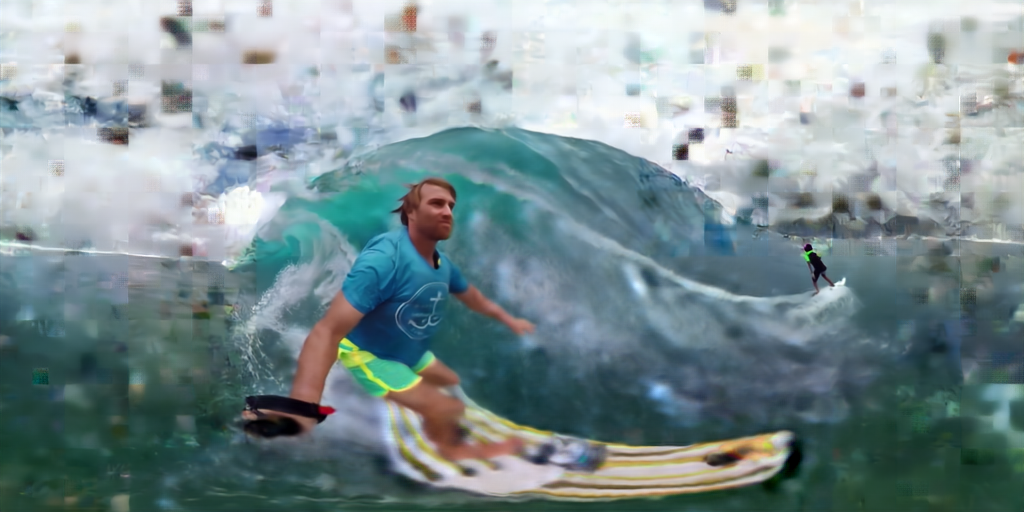} & 
        \includegraphics[width=0.23\textwidth]{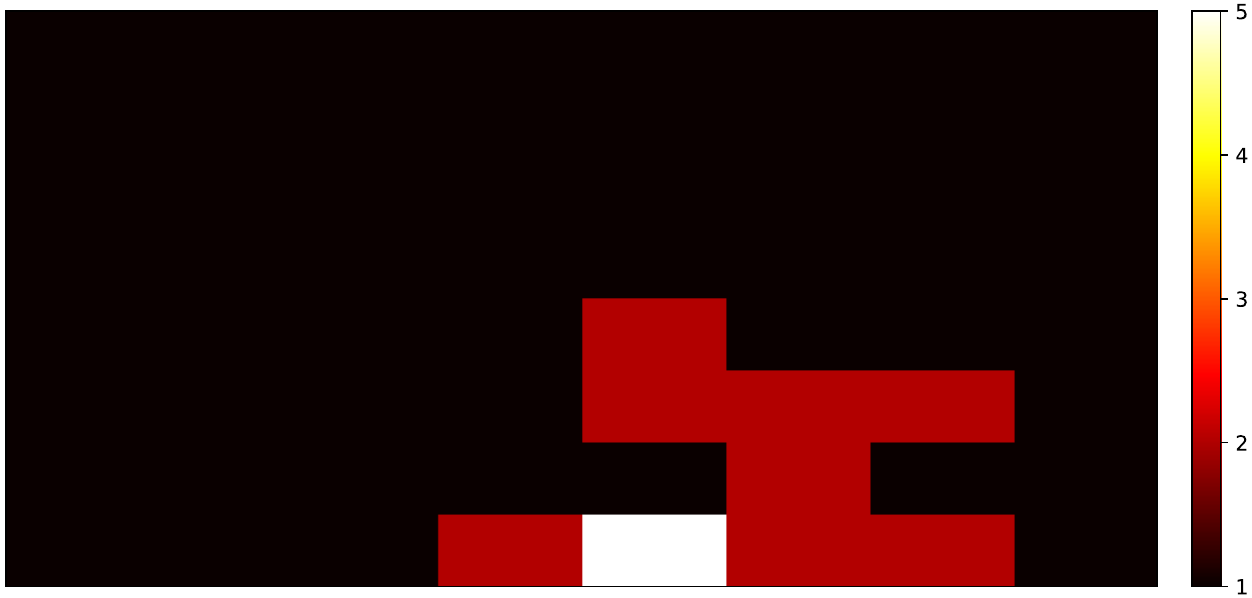}\\
        & CBR / WS-PSNR (dB) & 0.0129 / 20.902 & 0.0203 / 20.567 & \\
        frame 4 & \includegraphics[width=0.21\textwidth]{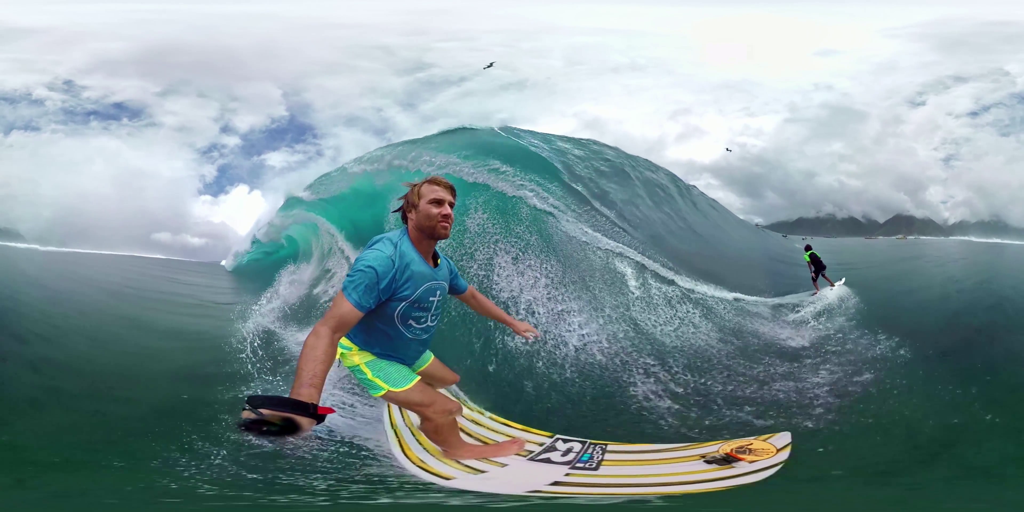} & 
        \includegraphics[width=0.21\textwidth]{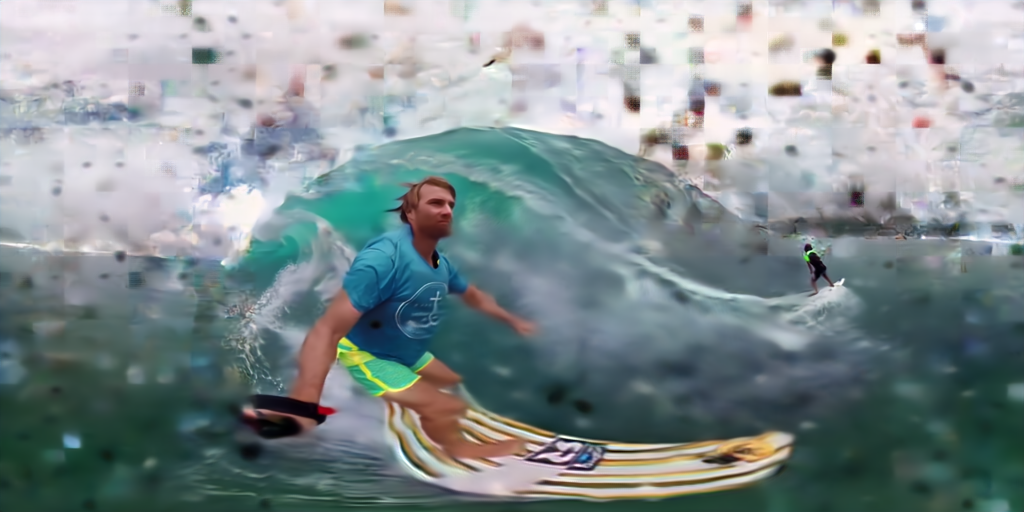} & 
        \includegraphics[width=0.21\textwidth]{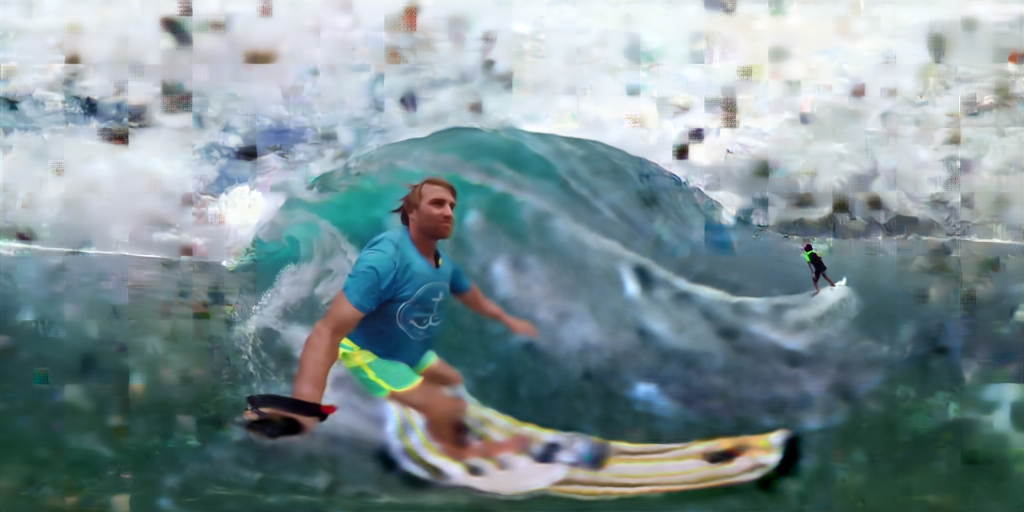} & 
        \includegraphics[width=0.23\textwidth]{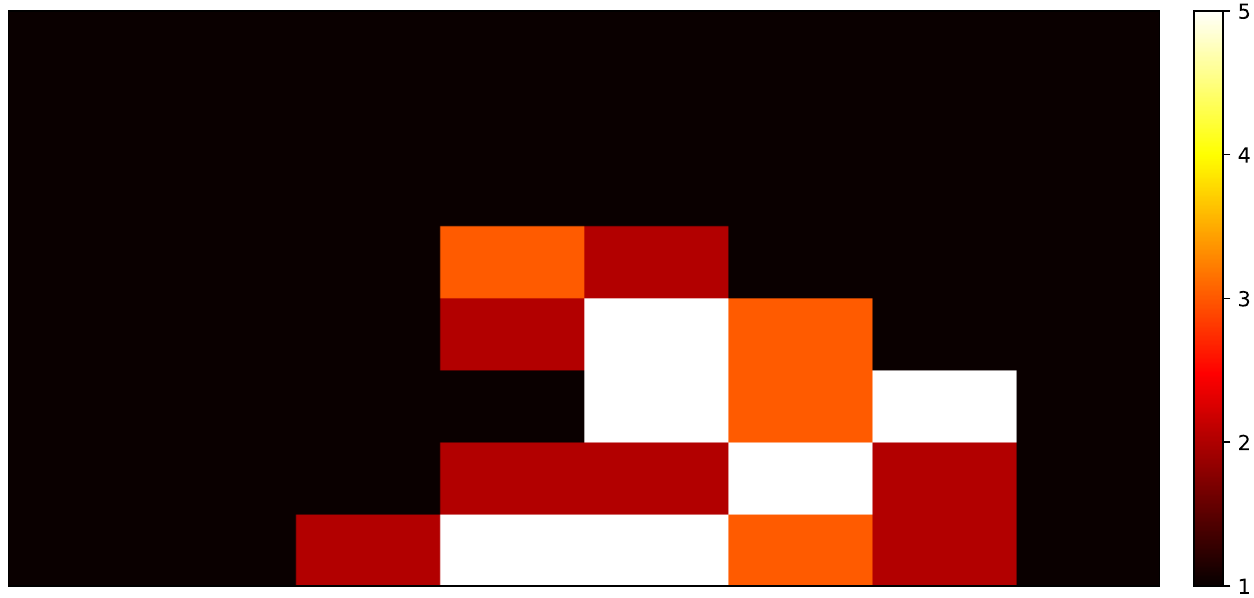}\\
        & CBR / WS-PSNR (dB) & 0.0131 / 20.390 & 0.0196 / 20.064 & \\
    \end{tabularx}
    \end{center}
    \caption{The reconstructed visualization of panoramic video cross-layer transmission, where the test SNR is 2dB. The first column shows the index of continuous frames in a panoramic video. The second column shows the original panoramic frame. The third and fourth columns show the reconstructed panoramic frame by APVST and DVST cross-layer transmission scheme, respectively. The fifth column shows the importance indication levels of the current frame transmission. Note that traditional communication schemes are completely incapable of achieving information recovery.}
    \vspace{-2mm}
    \label{cross-layer design visualization1}
\end{figure*}

\subsubsection{WS-SSIM Performance}
Fig.~\ref{wsssim vs others} shows the WS-SSIM achieved by APVST, APVST (w/o WA), DVST, H.264, and H.265 cross-layer transmission schemes versus CBR, SNR, and average retransmission times. During the training of the semantic network, the reconstruction loss is chosen as the negative of WS-SSIM. Similar to Fig.~\ref{wspnsr vs others}, experimental results show that semantic schemes can demonstrate superior transmission performance in cross-layer transfer with minimal resource consumption. As shown in Fig.~\ref{semantic wsssim vs cbr}, in achieving the same WS-SSIM, the proposed APVST can reduce bandwidth consumption by 44\% and 85\% compared to H.264 and H.265, respectively, in panoramic video cross-layer transmission. However, it can be observed that at lower CBRs, the WS-PSNR achieved by APVST, APVST (w/o WA), and DVST are quite similar. This indicates that the entropy calculated under the guidance of the entropy model and the latitude adaptive network is smaller, and consequently, the impact of $\mathcal{L}^\prime{_t^\mathrm{la}}$ in Eq.~\eqref{loss of la} is also reduced. Fig.~\ref{semantic wsssim vs snr} and Fig.~\ref{semantic wsssim vs re} show that in situations of poor communication quality or scarce communication resources, semantic schemes can still ensure the quality of information transmission by more effectively managing and utilizing limited resources.

\subsubsection{LPIPS Performance}
Fig.~\ref{lpips vs others} shows the LPIPS achieved by APVST, APVST (w/o WA), DVST, H.264, and H.265 cross-layer transmission schemes versus CBR, SNR, and average retransmission times. In addition to the previously mentioned WS-PSNR and WS-SSIM distortion metrics, aiming at the goals of semantic communications, the reconstruction loss chosen for training the semantic network is the human-perception-oriented LPIPS. As shown in Fig.~\ref{semantic lpips vs cbr}, at higher CBRs, the semantic scheme can better capture image details and textures, providing a more realistic and vivid visual experience to users. Additionally, Fig.~\ref{semantic lpips vs snr} shows that in extremely low SNR environments, traditional schemes degrade rapidly, whereas the semantic scheme, focusing more on human perceptual quality, can effectively resist noise interference, maintaining higher video quality. These simulation results indicate the compatibility and adaptation of the semantic scheme with traditional mobile communication systems, thus further enhancing the information transmission efficiency.

\subsubsection{The Results of Visualization}
We presented the visualization results of reconstructed frames under different schemes for cross-layer transmission of panoramic videos. As shown in Fig.~\ref{cross-layer design visualization1}, the first frame within a GoP has no reference frames, thus it is encoded and transmitted using the NTSCC. During this process, the entropy model in NTSCC identifies background information as a lower priority, while color-rich objects are recognized as a higher priority. Subsequent frames are transmitted using APVST, where its entropy model prioritizes regions of the video with relative motion as a higher priority, and relatively static regions are transmitted as a lower priority. Since 2ed and 4th frames are reconstructed based on their respective preceding frames, and the semantic scheme’s cross-layer transmission is inherently lossy, the clarity of the frames gradually decreases with the frame index, leading to a gradual decrease in WS-PSNR.

In the designed cross-layer transmission scheme, due to the limitations of transmission resources and the randomness of channel quality, the system prioritizes the transmission of higher priority information, resulting in clearer and more perceptually aligned parts. Conversely, lower priority parts are allocated fewer communication resources, leading to poorer quality in the recovered parts, a fact confirmed by simulation results. This also demonstrates the compatibility and adaptation of semantic communication with traditional communication systems, as well as the superiority of transmission performance.

\subsubsection{E2E Distance Effects}
\begin{figure}[ht]
\centering
\includegraphics[width=80mm]{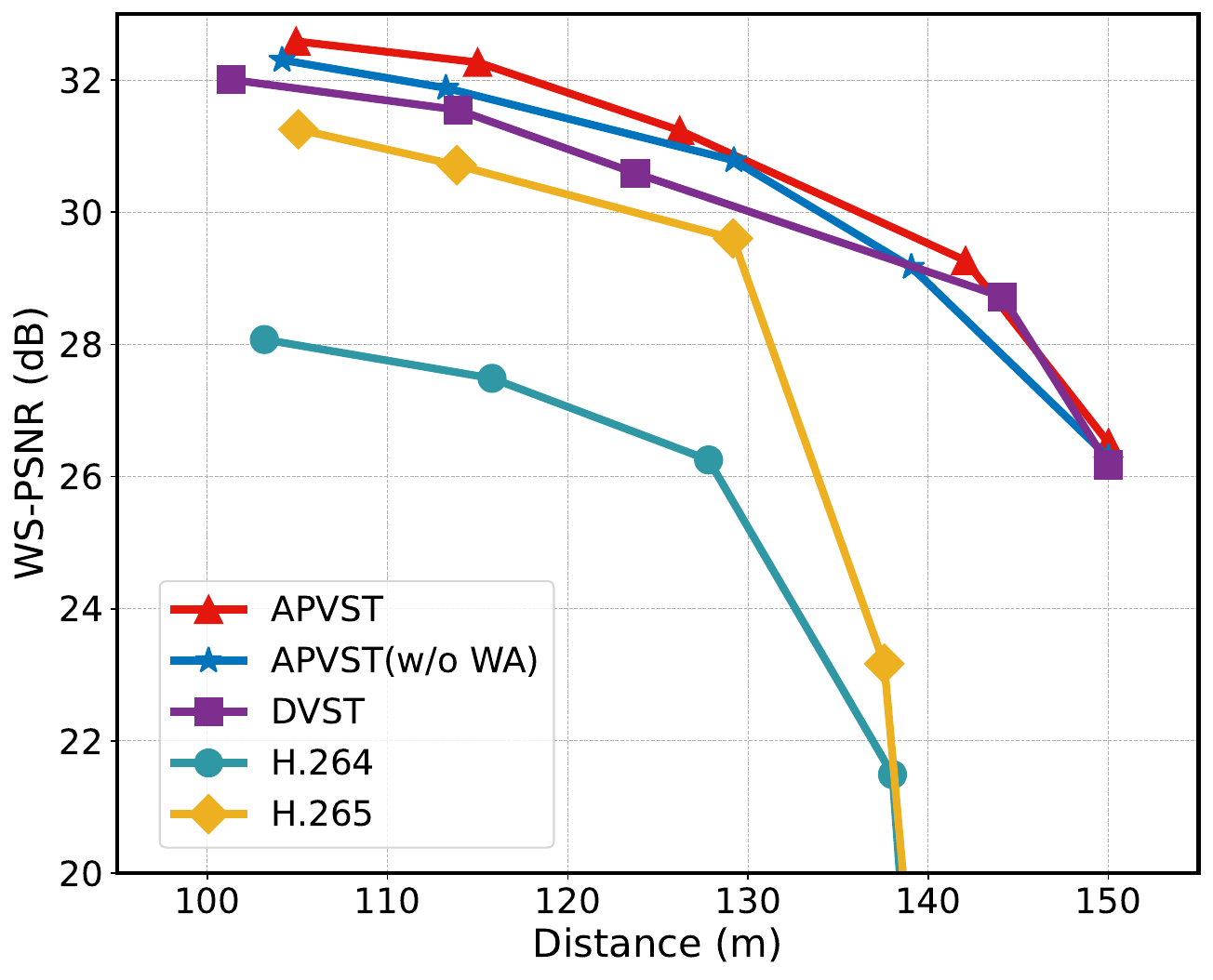}
\caption{WS-PSNR performance vs. the distance between the transmitter and the receiver at CBR=0.04 with BPSK modulation.}
\label{wspsnr vs distance}
\vspace{-4mm}
\end{figure}

Fig.~\ref{wspsnr vs distance} shows the WS-PSNR achieved by APVST, APVST (w/o), DVST, H.264, and H.265 cross-layer transmission schemes versus the distance between the transmitter and the receiver. As expressed in Eq.~\eqref{cal snr}, large-scale fading is directly affected by the distance, which in turn impacts the SNR. At closer distances, all schemes demonstrate commendable performance. However, as the distance increases, the fading caused by obstructions and multipath effects becomes more pronounced. Under these circumstances, traditional schemes exhibit a significant ``cliff effect'', where performance abruptly deteriorates. In contrast, semantic schemes maintain higher performance in cross-layer transmission. These results further substantiate the superior performance of our proposed CLESC scheme in long-distance transmissions.

\section{Conclusion} \label{conclusion}
This paper introduces a cross-layer encrypted semantic communication framework, i.e., CLESC. Compared to other semantic communication systems, the proposed CLESC framework incorporates semantic feature extraction and Deep JSCC and integrates encryption, CRC, and retransmission into the semantic information. This ensures compatibility and adaptation with traditional communication systems at low bandwidth cost. Concurrently, the CLESC framework exploits an adaptive cross-layer transmission mechanism, dynamically adjusting CRC, channel encoding, and retransmission schemes to prioritize important information. This approach ensures that critical data receives higher transmission priority, maintaining effective processing even under poor conditions. To verify the efficiency of CLESC, we take panoramic video transmission as an example and design an APVST scheme as an important part of the CLESC framework. APVST is based on a Deep JSCC structure and attention mechanisms and integrates the latitude adaptive module and weighted attention module to achieve more refined rate control and a better quality of users' immersive experiences. Extensive simulation results show that the proposed CLESC achieves compatibility and adaptation between semantic communication and traditional communication systems, enhancing transmission efficiency and effectively preventing the ``cliff effect'' associated with traditional communications. In the future, we will combine multimodal information to build a more comprehensive cross-layer compatible transmission framework to achieve more flexible data processing and transmission capabilities.

\end{document}